\documentclass[aps,prc,reprint,nofootinbib,longbibliography]{revtex4-2}
\usepackage[utf8]{inputenc}
\usepackage{graphicx,fancybox,float,comment}

\usepackage{units}
\usepackage{multirow,array}
\usepackage[dvipsnames]{xcolor}
\usepackage{amsmath,amssymb,amsfonts}

\usepackage{yfonts}
\usepackage{tensor}
\usepackage{paralist}   
\usepackage{braket}
\usepackage{tensor,mathrsfs}
\usepackage[normalem]{ulem}
\usepackage{hyperref}
\usepackage{cleveref}

\newcommand\ui[1]{\,\unit{#1}}

\begin{document}

\title{
Energy dependence of the chiral magnetic effect in expanding holographic plasma 
}

\newcommand{\uaAff}{\affiliation{Department of Physics and Astronomy, University of Alabama, 514 University Boulevard, Tuscaloosa, AL 35487, USA}}
\newcommand{\BNLAff}{\affiliation{Physics Department, Brookhaven National Laboratory, Upton, NY 11973, USA}}

\author{Casey Cartwright,}
\email{cccartwright@crimson.ua.edu} \uaAff
\author{Matthias Kaminski,}
\email{mski@ua.edu} \uaAff
\author{Bj{\"o}rn Schenke}
\email{bschenke@bnl.gov}\BNLAff

\date{\today}

\newcommand{\nm}{\nonumber}
\newcommand{\exd}{\mathrm{d}}
\newcommand{\tr}{\text{Tr}}
\newcommand{\zs}{z_*}
\newcommand{\cfour}{Q}
\newcommand{\vev}[1]{\braket{#1}}

\begin{abstract}
Based on a holographic far-from-equilibrium calculation of the chiral magnetic effect~(CME) in an expanding quark gluon plasma, we study collisions at various energies. 
We compute the time evolution of the CME current in the presence of a time-dependent axial charge density and subject to a time-dependent magnetic field. 
The plasma expansion leads to a dilution 
of the CME current. 
We study distinct combinations of how the initial magnetic field and initial axial charge behave with changing initial energy as proposed in previous literature. 
Most scenarios we consider lead to an increasing time-integrated CME current, when increasing the initial energy. This would make it more likely to observe the CME at higher collision energies. 
\end{abstract}

\maketitle
\tableofcontents

\section{Introduction}
\label{sec:introduction}
Significant experimental effort has been directed at the observation of the chiral magnetic effect (CME), which refers to the separation of electromagnetic charges along a magnetic field 
$\vec{B}$ \cite{Kharzeev:2004ey,Kharzeev:2007jp,Son:2009tf,Gynther:2010ed,Amado:2011zx,Kharzeev:2012ph,Ammon:2020rvg,Ammon:2017ded}, a signal of the presence of the chiral anomaly \cite{Bell:1969ts,Adler:1968av}.
The resulting charge current is referred to as {\it CME current}, which near equilibrium is given by~\cite{Neiman:2010zi,Son:2009tf,Kharzeev:2007jp}
\begin{equation}\label{eq:axialcurrent}
    \langle \vec{J}\rangle \propto C   \mu_5 \vec{B}  \, ,
\end{equation}
where the chiral anomaly coefficient $C$ can be computed from a triangle diagram~\cite{Bell:1969ts,Adler:1968av}, and $\mu_5$ is the axial chemical potential. 

The CME current has been shown to exist in the condensed matter physics context~\cite{Son:2012bg,Chernodub:2013kya,Cortijo:2015hlt,Landsteiner:2015lsa,Cortijo:2016wnf}, and it was first measured in condensed matter experiments~\cite{Li:2014bha,Arnold:2015vvs,Huang:2015,Xiong:2015,Li:2015,Li:2016,Hirschberger:2016,Gooth:2017mbd,Shekhar:2018}.  
After more than a decade of experimental searches in heavy ion collisions at both the Relativistic Heavy Ion Collider (RHIC) \cite{STAR:2009tro,STAR:2009wot,STAR:2013ksd,STAR:2013zgu,STAR:2014uiw,STAR:2019xzd,STAR:2021mii}, and the Large Hadron Collider (LHC) by the ALICE \cite{ALICE:2012nhw,ALICE:2017sss,ALICE:2020siw} and CMS \cite{CMS:2016wfo,CMS:2017lrw} Collaborations, a dedicated experiment was designed at RHIC aimed at observing a clean CME current. This experiment involved collisions of isobar nuclei, Zr+Zr and Ru+Ru collision systems, because one expected little difference in the overall collision geometries, but a larger magnetic field, and consequently a larger CME current, in Ru+Ru collisions, due to the larger electric charge of Ru versus Zr. 

Based on the data collected in this isobar run at RHIC~\cite{Marr:IPAC2019-MOZPLS2}, the STAR Collaboration recently reported that, based on the
{\it pre-blind criteria} which had been defined before the blind analysis, at 200 GeV nucleon-nucleon center of mass energy there was no signal of the CME~\cite{STAR:2021mii}.
The analysis by the STAR Collaboration~\cite{STAR:2021mii} was highly impressive and diligent. In the aftermath of their report it has become clear that Zr and Ru are not as similar as one had anticipated, in particular the differences in the collision geometry were underappreciated. Such differences had not been taken into account when constructing the criteria for a positive signal during the blind analysis~\cite{STAR:2021mii}. As a result, an updated analysis could lead to there being room for a CME current in the isobar data. However, due to the extreme complexity of the systems under investigation, there remain uncertainties, which at least in part can be decreased by a more accurate theoretical understanding. 

While the number of uncertainties is large, there are a few which are of primary importance to understanding CME physics in the isobar run. As mentioned above, there is an {\it uncertainty about the initial state} in each collision, i.e., the shape of the colliding ions as well as the individual distributions of protons and neutrons. Differences in the collision geometry can affect the definition of centrality, leading to differences in the charged hadron multiplicities between Ru+Ru and Zr+Zr systems, in the same centrality percentiles.
In addition, the charge distribution inside the nuclei will also affect the magnetic field generated by the colliding ions. This constitutes another known uncertainty: the {\it magnitude, spatial extent and time-evolution of the magnetic field over the lifetime of the plasma}, for which there exist many different (sometimes contradictory) predictions~\cite{Deng:2012pc,McLerran:2013hla,Roy:2015kma,Pu:2016ayh,Stewart:2017zsu,Oliva:2019kin,Yan:2021zjc,Stewart:2021mjz}. We will discuss this in more detail in the discussion, section~\ref{sec:discussion}.  

These uncertainties are often compounded by the need for time-evolution through (magneto)hydrodynamics or other approximate methods.\footnote{Also uncertainties in the hadronic phase and freeze-out/conversion to observable particles need to be considered.} Here, not only the regime of applicability of the method of time-evolution is uncertain, but so are, in addition, the necessary input parameters such as the {\it value of the axial charge density}. 
In particular it is not well understood how these parameters vary for collisions at different energies, {\it i.e., their beam energy dependence}. 
Early reports by the STAR collaboration~\cite{STAR:2014uiw} provided some information on how the CME current depends on the beam energy. However it is still unclear how much of the reported data in~\cite{STAR:2014uiw} can be attributed to the CME itself and how much was due to the background. 

The existence of these uncertainties provides ample motivation for us to further investigate the behavior of the CME in controlled settings. While we will not address issues of collision geometries, multiplicity distributions, or the distribution of the charges within the nuclei generating the magnetic field, we will present a holographic model in which the time-evolution of the axial charge and the magnetic field is governed by the symmetries of the system, manifest in the equations of motion. 

Various computations based on magnetohydrodynamics at late times support that during the collision the magnetic field first decays in vacuum and then more slowly, inversely proportional to the proper time, $B\propto \tau^{-1}$~\cite{Yan:2021zjc,Deng:2012pc}, as also expected from analytic solutions to Bjorken-like flow with magnetic field~\cite{Roy:2015kma,Pu:2016ayh}. 
In Bjorken-like flow, the axial charge density is also expected to decay inversely proportional to the proper time, $\vev{J^0_{(5)}}\propto \tau^{-1}$. 
From linearized hydrodynamics it is known that the CME current is proportional to the axial chemical potential $\mu_5$ and the magnetic field $B$, so that this Bjorken-like scenario would suggest a CME current proportional to $\tau^{-2}$. Then the question is if this current starts out large enough during the plasma phase to be detectable in the produced particles. 
Within our model we are going to answer this question.\footnote{ 
A magnetohydrodynamic computation of the effect of the CME in Bjorken-like flow was conducted in~\cite{Siddique:2019gqh} and its effect on the electromagnetic fields was described.
}

In this work, we address a subset of the uncertainties mentioned above. We start our holographic model evolution in a {\it known initial state} determined by the \textit{strongly coupled} field theory itself. We merely fix the initial value of the magnetic field, the initial value of the axial charge density, and require the electric field to vanish initially. 

As dictated by the holographic equations of motion, the {\it axial charge density decreases} as $ \vev{J^0_{(5)}}\propto \tau^{-1}$, while the \textit{magnetic field} decreases as $B\propto\tau^{-1}$, like in magnetohydrodynamic models~\cite{Yan:2021zjc,Roy:2015kma,Pu:2016ayh,Deng:2012pc}.
Our time-evolution includes the full \textit{far from equilibrium dynamics} which need not be hydrodynamic and need not be Bjorken-like.  

A similar holographic system was studied previously, it featured a non-expanding plasma, isotropic in the transverse plane and included a time-independent axial charge density and time-independent magnetic field~\cite{Ghosh:2021naw}. 
There, the authors find that at RHIC energies a non-expanding plasma can yield a significant CME current, while it does not at LHC energies. 
Here, considering several distinct scenarios, we indeed come to a different conclusion for most of those scenarios, as will be discussed in the remainder of this paper.  

In figure~\ref{fig:charge_totals}, we display the main result of our work, the charge accumulated during the expansion of the plasma in six scenarios (cases). Four out of those six cases lead to a larger accumulated charge for larger initial collision energies. 

We extend~\cite{Ghosh:2021naw} to include the effect of the expansion of the plasma, include the time-dependence for the axial charge density and magnetic field, start with an initial state anisotropic in the transverse plane and consider the pressure anisotropies as observables.  
A comparison of our results and~\cite{Ghosh:2021naw} is given in section~\ref{sec:results}. 
In terms of the time-evolution of the plasma, the axial charge and the magnetic field,  our model plasma extends this non-expanding case~\cite{Ghosh:2021naw} as well as the (ideal) magnetohydrodynamics Bjorken-expanding case, which is discussed in~\cite{Siddique:2019gqh}. 

In section~\ref{sec:model}, we introduce the gravitational holographic model, present our numerical results for the CME current and pressure anisotropies in section~\ref{sec:results}, and conclude with the discussion section~\ref{sec:discussion}. Appendix~\ref{append:EOM} contains the gravitational equations of motion and appendix~\ref{append:asymp} details the near-boundary expansions of each field in the gravity theory. Appendix~\ref{append:num} briefly details the numerical methods we use to solve the time-dependent Einstein equations.

\section{Holographic Model}
\label{sec:model}
In this section we construct a holographic model which at late times follows Bjorken expansion along the ``beamline" ($x_3$-direction), which is subject to a magnetic field along one transverse direction ($x_1$-direction), and which contains a nonzero axial charge density. Both, the magnetic field and the axial charge density are time-dependent. 

\subsection{Gravitational theory}
Consider the five-dimensional gravity action~\cite{Gynther:2010ed,Ghosh:2021naw}\footnote{In our conventions, Greek indices are used for 5-dimensional coordinates $x^\mu,x^\nu, ...= \{r, v, x_1, x_2, \xi\}$, and Latin indices for the 4-dimensional field theory coordinates $x^a,x^b, ... =\{ v, x_1, x_2, \xi \} $, and the second part of the Latin alphabet for 3-dimensional spatial directions $x^i,x^j,... = x_1,x_2,\xi$. }
\begin{align}\label{eq:action}
    S &= \frac{1}{16\pi G} \int \exd^5 x \sqrt{-g} 
    \left[ 
         R-2\Lambda  
    \right. \nonumber\\
    & \left.   -\frac{L^2}{4}
     \left( 
       F_{\mu\nu}F^{\mu\nu} + F^{(5)}_{\mu\nu}F_{(5)}^{\mu\nu} 
     \right) 
     \right. \nonumber\\
    & \left. 
    +\frac{\alpha}{3}\epsilon^{\mu\nu\rho\sigma\tau}A_{\mu}
    \left(
        3F_{\nu\rho}F_{\sigma\tau}+F^{(5)}_{\nu\rho}F^{(5)}_{\sigma\tau}
    \right)
    \right]+S_{ct} \, ,
\end{align}
whose holographic dual enjoys a global $U(1)_A\times U(1)_V$ symmetry. Here, $R$ is the Ricci scalar and $g$ is the determinant of the five-dimensional metric $g_{\mu\nu}$. $G$ is the five-dimensional Newton constant, $\Lambda$ is the cosmological constant, which in AdS$_{4+1}$ is given in terms of the radius of AdS, $L$, namely $\Lambda=-6/L^2$, and $\alpha$ is the Chern-Simons coupling. The gauge fields associated with this vector and axial $U(1)$-symmetry in the bulk theory are given as $A^{\mu}$ and $V^{\mu}$, respectively, with their field strengths\footnote{The position of the subscript (or superscript) $(5)$ is arbitrary and serves only to distinguish the axial field strength from its vector companion.} defined as $F^{(5)}_{\mu\nu}=\partial_{[\mu,} A_{\nu]}$ and $F_{\mu\nu}=\partial_{[\mu,} V_{\nu]}$ where we have made use of the square brackets to indicate an antisymmetrization of the indices, {i.e.},  $A_{[\mu}B_{\nu]}=A_\mu B_\nu-A_\nu B_\mu$.
The Levi-Civita tensor in this work is defined in terms of the totally anti-symmetric Levi-Civita symbol $\epsilon(\mu\nu\rho\sigma\tau)$ as $\epsilon^{\mu\nu\rho\sigma\tau}=\epsilon(\mu\nu\rho\sigma\tau)/\sqrt{-g}$. 

The counter-term action, $S_{ct}$, contains the relevant counter terms required to render the variational problem well posed (here presented for a $(d+1)$-dimensional bulk spacetime; $d=4$ in our case)~\cite{Taylor:2000xw}
\begin{align}
    S_{ct}&=\frac{1}{8\pi G}\int\exd^d x\sqrt{\gamma}\left( K -\frac{1}{2L}\left(2(1-d) \right.\right. \nonumber\\
    &\left.\left. -\frac{L^2}{d-2}R(\gamma)  \right)\right)+\frac{L^3}{64\pi G}\log(\epsilon)\int\exd^d x\sqrt{\gamma_0}F_0^2\, ,\label{eq:counter_term_action}
\end{align}
where $K$ is the trace of the extrinsic curvature, 
$\gamma$ is the induced metric on a constant $r=1/\epsilon$ hypersurface (with a small cut-off energy value $\epsilon$), $\gamma_0$ is the metric of the dual field theory and $F_0$ is the external field strength of the gauge field $V$ in the dual theory\footnote{We do not include the dual axial field strength, $F^{(5)}_0$, in eq.~(\ref{eq:counter_term_action}) since it vanishes for the ansatz we consider.}. 

The equations of motion, which follow from the action are given by,
\begin{subequations}
\begin{align}
&R_{\mu\nu}-\frac{1}{2}g_{\mu\nu}(R-2\Lambda)=\frac{L^2}{2}T_{\mu\nu}+\frac{L^2}{2}T^{(5)}_{\mu\nu} \, , \label{eq:eom_E}\\
&\nabla_{\mu}F^{\mu\nu} =-\frac{2\alpha}{L^2} \epsilon^{\nu\beta\lambda\rho\sigma} F_{\beta\lambda} F^{(5)}_{ \rho\sigma} \, , \label{eq:eom_V}\\
&\nabla_{\mu}F_{(5)}^{\mu\nu} =-\frac{\alpha}{L^2} \epsilon^{\nu\beta\lambda\rho\sigma}\left( F_{\beta\lambda} F_{ \rho\sigma} +F^{(5)}_{\beta\lambda} F^{(5)}_{ \rho\sigma}\right) \, , \label{eq:eom_A}
\end{align}
\end{subequations}
where we have defined the standard energy-momentum tensor associated with the gauge field Lagrangian,
\begin{equation}
    T_{\mu\nu}=F_{\mu\lambda}\tensor{F}{_\nu^\lambda}-\frac{1}{4}g_{\mu\nu}F^2,\, \,  T^{(5)}_{\mu\nu}=F^{(5)}_{\mu\lambda}\tensor{F}{^{(5)}_\nu^\lambda}-\frac{1}{4}g_{\mu\nu}F_{(5)}^2\, .
\end{equation}
and $R_{\mu\nu}$ is the Ricci tensor derived from the spacetime metric $g_{\mu\nu}$. We are interested in bulk solutions which at the asymptotic AdS$_5$ boundary are undergoing a Bjorken expansion at late times. 
In the bulk five dimensional space we work with generalized Eddington-Finkelstein coordinates  $(r,v,x_1,x_2,\xi)$, adapted to reduce to Milne coordinates $(\tau,x_1,x_2,\xi)$ at the AdS-boundary. Here $r$ is the radial AdS direction, $v$ the Eddington-Finkelstein time, $x_1,x_2$ coordinates in the transverse plane, $\xi=\frac{1}{2} \ln [(t+x_3)/(t-x_3)]$ is the spacetime rapidity and $\tau=\sqrt{t^2-x_3^2}$ is the proper time for which 
\begin{equation}
    \tau=\lim_{r\rightarrow \infty}v\, .
\end{equation} 
That is, the bulk Eddington-Finkelstein time $v$ reduces to the Milne proper time at the AdS-boundary. 

\subsection{Metric and gauge field ansatz}\label{sec:ansatz}
As an ansatz for the bulk five-dimensional metric we take\footnote{This is similar to the ansatz used in~\cite{Chesler:2008hg} by the authors who originally used this method in asymptotically $AdS$ spacetimes and is an extension of the ansatz used in~\cite{Cartwright:2020qov} to study the time evolution of the axial current in an explicitly top-down holographic model. 
}
\begin{align}\label{eq:ansatz_metric}
    ds^2&=2 \exd r\exd v -A(v,r)\exd v^2+F_1(v,r)\exd v\exd x_1
     \nonumber \\
    &+S(v,r)^2e^{H_1(v,r)}\exd x_1^2+S(v,r)^2e^{H_2(v,r)}\exd x_2^2 \nonumber\\
    &+ L^2S(v,r)^2e^{-H_1(v,r)-H_2(v,r)}\exd \xi^2,
\end{align}
where $A,F_1,S,H_1$ and $H_2$ are scalar functions of the bulk radial coordinate $r$ and the Eddington-Finkelstein time $v$. To obtain an expanding $(3+1)$-dimensional spacetime at the conformal boundary of the $AdS_{4+1}$ spacetime, we place the following boundary condition on the metric at the conformal boundary
\begin{equation}
    \lim_{r\rightarrow \infty} \frac{L^2}{r^2}\exd s^2=-\exd \tau^2+\exd x_1^2+\exd x_2^2+\tau^2\exd \xi^2 \, .\label{eq:nearbdy}
\end{equation}
In terms of the components of the metric, the boundary condition in eq.\,(\ref{eq:nearbdy}) implies 
\begin{align}
  \lim_{r\rightarrow \infty}  A&\rightarrow \left(\frac{r}{L}\right)^2 \, , \\
  \lim_{r\rightarrow \infty}  H_1&\rightarrow -\frac{2}{3} \log \left(\frac{\tau}{L}\right)\, , \\
  \lim_{r\rightarrow \infty}  H_2&\rightarrow -\frac{2}{3}\log \left(\frac{\tau}{L}\right) \, , \\
  \lim_{r\rightarrow \infty}    S&\rightarrow r \left(\frac{\tau}{L^4}\right)^{1/3} \, , \\
  \lim_{r\rightarrow \infty}    F_1&\rightarrow 0 \, .
\end{align}

As an ansatz for the gauge fields we take
\begin{align} \label{eq:ansatz_gauge}
    A_{\mu}&=\frac{1}{L}(0,-\phi(v,r),0,0,0), \nonumber \\ 
    V_{\mu}&=\frac{1}{L}(0,0,-V(v,r),b\, \xi,0)\, ,
\end{align}
with the dimensionless magnetic field, $b$. Note, that we choose the ordering $(r,v,x_1,x_2,\xi)$ when writing vector components explicitly as in eq.~\eqref{eq:ansatz_gauge}. 

Considering the field strength associated with $V_\mu$, we see that our ansatz produces a vector magnetic field directed along the $x_1$-direction. This vector magnetic field will persist in the field theory at the conformal boundary of our spacetime. We can compute the precise form of the dual external gauge field in the boundary theory by considering,
\begin{equation}
    \lim_{r\rightarrow \infty} V_{a} = V_{a}^{\text{ext}}=\frac{1}{L}\left(0,0,b \, \xi,0\right) \, , \label{eq:ext}
\end{equation}
This implies that the only nonzero components of the dual field strength are $F_{\xi 2}=-F_{2\xi}=b/L$. 
Given the fluid velocity, $u^{a}$, the magnetic field as seen in a fluid cell is,
\begin{equation}
    B^a=\frac{1}{2}\epsilon^{abcd}u_bF_{cd}\quad \Rightarrow\quad B^1=\frac{b}{L \tau} \, .
\end{equation}
where we have used the fluid velocity in the co-moving frame at the boundary $u^{a}=(1,0,0,0)$. 
The choice of the other components of the gauge fields are motivated by our desire to capture essential chiral magnetic physics. The bulk quantity $V(v,r)$ will soon be shown to produce\footnote{Note that in eq.~(\ref{eq:ext}) we have implicitly chosen the source of $V_1$ to be zero. i.e. $V_1\sim \mathcal{O}(r^{-2})$ near the conformal boundary. This choice implies that we can consistently set the metric function $F_1=0$. If we were to keep this term and provide a source for this component of the vector gauge field, the metric component $F_1$ would describe the backreaction of the CME current on the energy-momentum tensor~\cite{Cartwright:2020qov,Ghosh:2021naw}. } the $x_1$ component of the vacuum expectation value of a global $U(1)$ vector charge current operator in the dual field theory (see eq.\ (\ref{eq:dual_currents_V})). The temporal component of the axial gauge field will provide the axial chemical potential in the dual theory. 

Our choice of ansatz realizes a $SO(3)$-symmetry which has been entirely broken in a controlled way by the expansion along $x_3$ on one hand, and by the magnetic field along $x_1$ on the other hand. 
Inserting the ansatz from eq.\,(\ref{eq:ansatz_gauge}) and eq.\,(\ref{eq:ansatz_metric}) into eq.\,(\ref{eq:eom_V}) and eq.\,(\ref{eq:eom_A}), one obtains three equations. One of these equations is for the evolution of the bulk gauge field component $V$ (which is included with the Einstein equations during the evolution) while the other two are for the axial scalar potential $\phi$. The equations for the scalar potential can be solved~\cite{Ghosh:2021naw} once it is realized that they can be written as a total derivative and hence determine a constant of motion,
\begin{align}
    q_5/L&=L^4S(v,r)^3\phi'(v,r)+8 \alpha  b V(v,r) \, ,\\ \mathcal{E}_5&\equiv-\phi'(v,r)=\frac{q_5L^{-1}-8 \alpha  b V(v,r)L^{-4}}{S(v,r)^3} \, ,\label{eq:electric}
\end{align}
where $\mathcal{E}_5$ is the bulk electric field. Here we have used a prime, $\phi'$, to denote a derivative with respect to the radial coordinate and $q_5$ is a conserved quantity. We will show that $q_5$ is directly related to the axial charge density in the dual field theory in section~\ref{sec:Operator_Expectations}. The result in eq.\,(\ref{eq:electric}) is highly useful numerically. The energy-momentum source, which appears on the right hand side of the Einstein field equations, is composed of contractions of the vector and axial field strengths, which themselves are composed of derivatives of the vector and axial gauge fields. Hence it is the bulk electric field $\mathcal{E}_5$, the derivative of the temporal component of the axial gauge field, which explicitly appears in the Einstein field equations, as can be seen in appendix~\ref{append:EOM} in the appearance of the terms proportional to $q_5$. 

The equations of motion that follow from eq.\,(\ref{eq:eom_E}) can be rewritten into a pseudo-nested\footnote{We call this a pseudo-nested list since the nesting is only partial. In this case three of the equations must be solved as a coupled system while the coupled system itself sits within the nested structure.} list by the definition of an additional derivative denoted by $\dot{f}$. The definition of the dotted derivative is given by
\begin{equation}\label{eq:char_dot}
    \dot{f}\equiv\partial_v f+\frac{1}{2}A \partial_r f \, ,
\end{equation}
and constitutes a derivative along outgoing null geodesics~\cite{Chesler:2013lia}. The pseudo-nested list generated by treating the directional derivatives of the metric components along outgoing geodesics as auxiliary fields is collected for this system in appendix~\ref{append:EOM}. The method of solution is directly analogous to earlier work~\cite{Cartwright:2020qov} although we do repeat it in appendix~\ref{append:num} for completeness.

\subsection{Energy-momentum tensor and currents of the dual boundary field theory}\label{sec:Operator_Expectations}
Using the standard holographic dictionary~\cite{Skenderis:2008dg,Taylor:2000xw,Fuini:2015hba,Pendas:2019} we can obtain the vacuum expectation values $\vev{\cdot}$ of the operators in the boundary corresponding to the energy-momentum tensor $T_{ab}$, the vector current $J^a$ and the axial current $J^a_{(5)}$. We will not go into detail as to how one obtains these results, instead we refer the interested reader to~\cite{Skenderis:2008dg,Taylor:2000xw} for more details on how one extracts this information using the holographic dictionary. For the vacuum expectation value of the energy-momentum tensor we find the following results: 
\begin{widetext}
\begin{align}
    \epsilon=\braket{T_{00}}&=\frac{2L^3}{\kappa^2}\left(-\frac{3 a_4(\tau)}{4L^4}-\frac{b^2 \log (b^{1/2} )}{8 L^2 \tau^2}\right) \ , \label{eq:HR_Energy}\\
    P_1= \braket{T_{11}}&=\frac{2L^3}{\kappa^2}\left( -\frac{a_4(\tau)}{4L^4}+\frac{h^{(1)}_4(\tau)}{L^4}+\frac{b^2 \log (b^{1/2} )}{8L^2 \tau^2}-\frac{1}{6 \tau^4} \right) \, , \label{eq:HR_p1}\\
    P_2= \braket{T_{22}}&=\frac{2L^3}{\kappa^2}\left(-\frac{a_4(\tau)}{4L^4}+\frac{h^{(2)}_4(\tau)}{L^4}-\frac{b^2 \log (b^{1/2} )}{8 L^2 \tau^2}-\frac{b^2}{16 L^2\tau^2}-\frac{1}{6 \tau^4} \right) \, ,\label{eq:HR_p2} \\
    \tau^2P_\xi= \braket{T_{\xi\xi}}&=\frac{2L^3\tau^2}{\kappa^2}\left(-\frac{a_4(\tau)}{4L^4}-\frac{h^{(1)}_4(\tau)}{L^4}-\frac{h^{(2)}_4(\tau)}{L^4}-\frac{b^2 \log (b^{1/2} )}{8 L^2\tau^2}-\frac{b^2}{16L^2 \tau^2}+\frac{1}{3 \tau^4}\right) \, ,\label{eq:HR_p3}
\end{align}
\end{widetext}
where $\epsilon$ is the energy density, and $P_1, P_2$ and $P_\xi$ are the pressures in the $x_1$-, $x_2$- and $\xi$-directions, respectively.  For the axial as well as the vector currents we obtain\footnote{Note, that these are the {\it consistent currents} (as opposed to the covariant currents)~\cite{Bardeen:1984pm,Landsteiner:2016led,Ammon:2020rvg}. With this definition the vector current is conserved and the axial current is equal to the consistent chiral anomaly.  
A standard Bardeen-Zumino term could be added to the generating function to produce the covariant currents.
}
\begin{align}
    \braket{J_{(5)}^{a}}&=\frac{1}{2\kappa^2}\left(\frac{q_5 L}{\tau},0,0,0\right), \label{eq:dual_currents_A} \\ \braket{J^{a}}&=\frac{1}{2\kappa^2}\left(0,2V_2(\tau),0,0\right)\, , \label{eq:dual_currents_V}
\end{align}
where the ordering of components is $(\tau,x_1,x_2,\xi )$. 
The $\tau$-dependence of the axial current~\eqref{eq:dual_currents_A} is due to the expansion along the $x_3$-direction. 
In equations~\eqref{eq:dual_currents_A}  and~\eqref{eq:dual_currents_V} we have used the standard definition of the gravitational constant $\kappa^2=8\pi G$. The non-vanishing components of the dual vector field strength are given by $F_{\xi 2}=-F_{2\xi}=B$ while the dual axial field strength is identically zero, $F^{(5)}=0$. 

The quantities that appear in the energy-momentum tensor as well as in the vector current, namely $a_4(\tau), h^{(1)}_4(\tau), h^{(2)}_4(\tau)$ and $V_2(\tau)$ are coefficients which result from an expansion of the $(4+1)$-dimensional metric ($g_{\mu\nu}$), and gauge fields ($A_{\mu}$ and $V_{\mu}$) near the conformal boundary. This expansion can be obtained by solving the Einstein equations order by order in the holographic coordinate $r$, near $r\rightarrow\infty$. These expansions are collected and displayed in appendix~\ref{append:asymp}. The goal of the numerical evolution is to extract these coefficients in order to construct with them the vacuum expectation values of the dual operators.  

Conformal symmetry is explicitly broken by the magnetic field, the trace of the energy-momentum tensor is proportional to the electromagnetic field strength as can be explicitly checked,
\begin{equation}
    \vev{T^a_a}=\frac{L^3}{8\kappa^2}F^2=-\frac{b^2 L}{4 \kappa ^2 \tau^2}=-\frac{B_1^2 L^3}{4 \kappa ^2}\, .
\end{equation}
One can see that at asymptotically late times conformal symmetry is restored as $\tau\to \infty$ while $b$, $\kappa$, and $L$ are constant over time. From eq.~(\ref{eq:dual_currents_A}) the axial charge density $\braket{J_{(5)}^{0}}$ is completely determined up to an initial constant, even out of equilibrium. This initial constant value was determined as the conserved quantity $q_5$ in section~\ref{sec:ansatz}. As expected from eq.~\eqref{eq:dual_currents_A} with the fact that $q_5$ is constant in $\tau$ and the Bjorken-expansion in one spatial direction, this current density falls off as $\tau^{-1}$. This fall-off will manifest itself in a finite time available for the CME vector current to grow before it is diluted due to the expansion of the medium. 

As an aside, we conclude this subsection by putting into context how the standard Navier-Stokes equations relate to the equations governing the time-evolution of the energy density in our holographic model. 
Of particular interest for comparison with known equations is the evolution equation for the energy density of the plasma. In (0+1)-dimensional Bjorken dynamics, the Navier-Stokes equations for an uncharged fluid reduce to a single equation for the energy density~\cite{Critelli:2018osu},
\begin{equation}
  \partial_\tau\epsilon+\frac{4}{3}\frac{\epsilon}{\tau}-\frac{4}{3}\frac{\eta}{\tau^2}=0\, .\label{eq:Navier}
\end{equation}
The equivalent equation in the system we evolve can be determined from a near-boundary solution to the Einstein equations as displayed in appendix~\ref{append:asymp}. Making use of the energy density and pressures as defined in eq.~(\ref{eq:HR_Energy}) through eq.~(\ref{eq:HR_p3}), the evolution equation is given by,
\begin{equation}
    -\frac{P_1(\tau)}{\tau }-\frac{P_2(\tau)}{\tau }-\frac{B_1(\tau)^2}{8 \tau }+\partial_\tau \epsilon(\tau) +\frac{2 \epsilon (\tau )}{\tau }=0 \, ,\label{eq:EQ_for_energy}
\end{equation}
where we recall $B_1(\tau)=\frac{b}{L\tau}$. In the late time limit one can show that solutions for the energy density as obtained from eq.~(\ref{eq:EQ_for_energy}) asymptote to the solutions of eq.~(\ref{eq:Navier}).

\subsection{Preparation of initial conditions} 
Initial conditions in this system consist of a choice of the following quantities: an initial time $\tau_0$, the coefficient $a_4(\tau_0)$ associated with the initial energy density, initial axial charge density $q_5(\tau_0)$, initial magnetic field $B_1(\tau_0)$ and a choice of the profile for the functions $H_1(v_0,z)$, $H_2(v_0,z)$, $V(v_0, z)$ along the (inverted) bulk AdS radial direction, $z=L^2/r$.
The choice of the functions $H_1(v_0,z)$, $H_2(v_0,z)$, $V(v_0, z)$ implicitly sets the initial pressure anisotropies as well as the initial value of the vector current in the dual theory. In all computations we set the initial, radial, profile for the Maxwell field $V$ in the vector $U(1)$ sector to zero, $V(v_0,z)=0$ and hence $\vev{J^1(\tau_0)}=0$ for every evolution. 
This means that each case which we study starts with a vanishing CME current at $\tau=\tau_0$. So, the current will grow for $\tau>\tau_0$ and will simultaneously be diluted as the plasma expands. 

We parametrize the initial profile for the metric components $H_i$ in terms of a deviation, $H_i^{(d)}$, away from the vacuum AdS solution, 
\begin{equation}
    H_i=H^{(\text{AdS})} +H^{(d)}_i\, , \quad H^{(\text{AdS})}=-\frac{2}{3}  \log \left(v+z\right) \, . \label{eq:init_Data}
\end{equation}
In every evolution we choose to set $H^{(d)}_1(v_0,z)=H^{(d)}_2(v_0,z)$. 
This choice implies a relation between asymptotic coefficients of the metric at the initial time, $h^{(1)}_4(\tau_0)=h^{(2)}_4(\tau_0)=h_4(\tau_0)$, and also implies the following about the pressure anisotropies at the initial time,
\begin{widetext}
\begin{align}
    \Delta P_{12}\equiv P_{1}-P_{2}&=\frac{2L^3}{\kappa^2}\left(\frac{b^2}{16 L^2 \tau_0 ^2}+\frac{b^2 \log (b)}{4 L^2 \tau_0 ^2}\right) \, ,\\
     \Delta P_{1\xi}\equiv P_{1}-P_{\xi}&=\frac{2L^3}{\kappa^2}\left(\frac{b^2}{16 L^2 \tau _0^2}+\frac{b^2 \log (b)}{4 L^2 \tau _0^2}+\frac{3 h_4\left(\tau _0\right)}{L^4}-\frac{1}{2 \tau _0^4}\right) \, , \\
 \Delta P_{2\xi}\equiv P_{2}-P_{\xi}&= \frac{2L^3}{\kappa^2}\left(\frac{3 h_4\left(\tau _0\right)}{L^4}-\frac{1}{2 \tau _0^4}\right) \, .
\end{align}
\end{widetext}
The parametrization of the initial data in eq.~(\ref{eq:init_Data}) must then be translated to the parametrization of the metric functions used in the numerical evolution scheme, where we work with scaled and subtracted metric components~\cite{Chesler:2013lia}, e.g.\ 
\begin{equation}
    H_i(z,v)=\Delta_{H_i}(z,v)+z^4 \tilde{H}_i(z,v) \, , \label{eq:numerical_data}
\end{equation}
where the scaling is determined so that the value of the subtracted function $\tilde{H}_i$ at the AdS-boundary is the unknown asymptotic coefficient of interest, e.g.\
\begin{equation}
    \lim_{z\rightarrow 0}\tilde{H}_i=h_4^{(i)} \, .
\end{equation}
The functions $\Delta_{H_i}$ represent singular and regular terms in the functions $H_i$. These functions are referred to as subtractions. We make use of these subtractions so that the function for which we solve numerically, the tilde quantity $\tilde{H}_i$, is a regular function. As a result, in practice, we never compute the ``un"-tilded functions. Rather our numerics make use of only tilded functions, as regular functions they can be represented with high accuracy by the Chebyshev basis we employ.

\begin{figure}[H]
    \centering
    \includegraphics[width=0.47\textwidth]{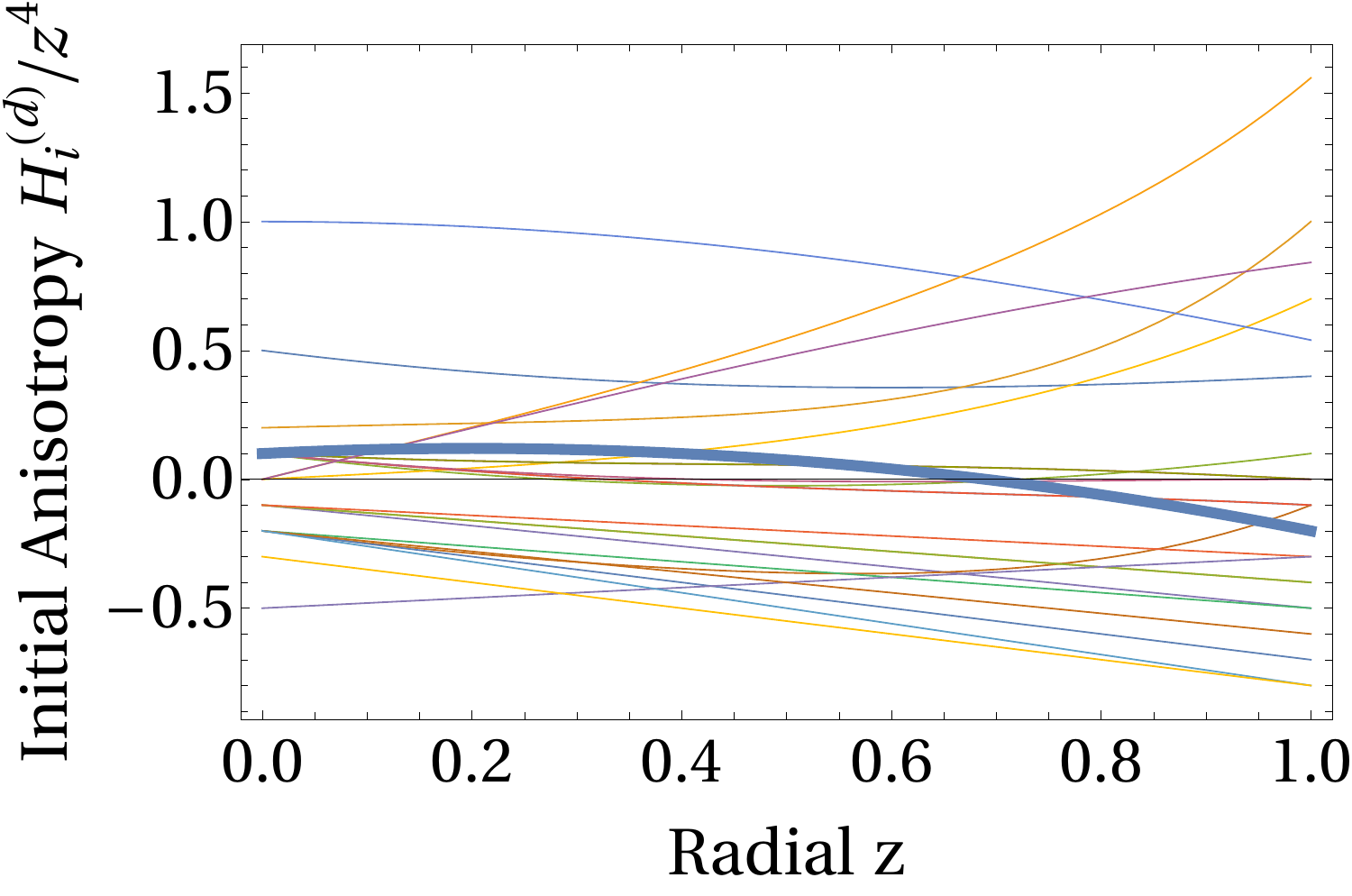} \hfill
    \includegraphics[width=0.47\textwidth]{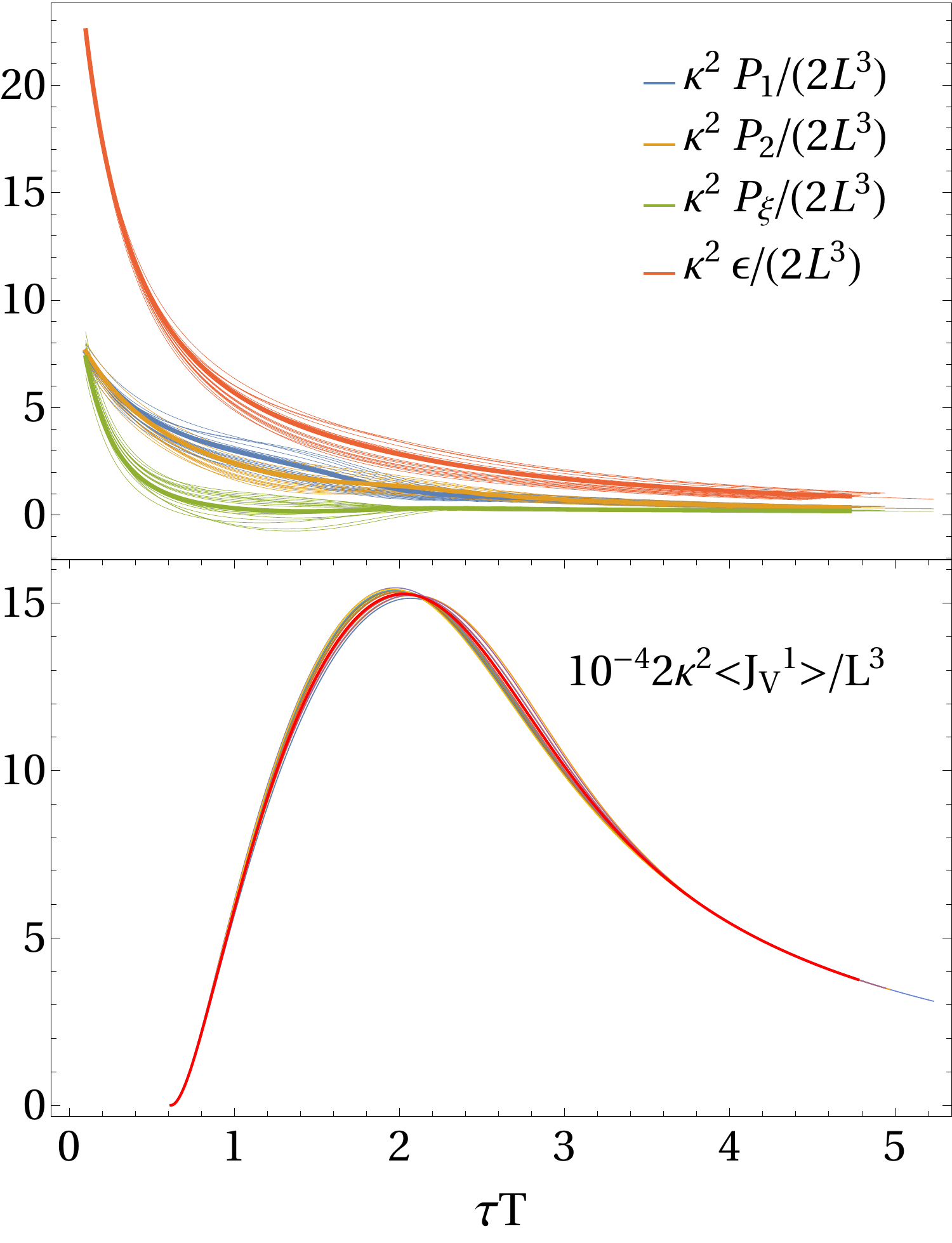}
    \caption{\textit{Top:} Deviation of the bulk spacetime anisotropy from the vacuum solution on the initial time slice. The thick curve displays the choice of initial data which we later investigate further in figure~\ref{fig:charge_totals} and figures~\ref{fig:vec_current}-\ref{fig:vec_current_charge_mag_scaling}. The thick line in the bottom two figures corresponds to results using the choice of parameters leading to thee thick line in the top plot. \textit{Middle:}  The corresponding evolution of the energy-momentum tensor components: energy density (\textcolor{red}{red}), and the three distinct pressures, along the beamline (\textcolor{green}{green}), and in the transverse plane along the magnetic field (\textcolor{blue}{blue}) and perpendicular to it (\textcolor{orange}{orange}). \textit{Bottom:} The vector current $J_V^1$, the CME current, along the magnetic field. 
    \label{fig:initial_data}}
\end{figure}

We take the following ansatz for the initial deviations from the AdS vacuum solution~\cite{Rougemont:2021qyk}
\begin{align}
    H^{(d)}_i(z)&=z^4 \Omega_1 \cos (\gamma_1 z)+z^4 \Omega_2 \tan (\gamma_2 z) \, ,\nonumber \\
    &+z^4 \Omega_3 \sin (\gamma_3 z) +z^4\sum_{j=0}^{4}\beta_j z^j\label{eq:init_data} \, ,
\end{align}
where $\gamma_i,\Omega_i,\beta_i$ are parameters which we choose to influence the behavior of the anisotropy functions in the bulk AdS spacetime. For the choice of parameters we use in this work the initial deviation of the bulk spacetime anisotropy from the vacuum solution is shown in the top panel of figure~\ref{fig:initial_data}. The values we choose for this work correspond to all 23 initial conditions\footnote{We do not use initial condition 12 and we have changed the value of one coefficient of initial condition 23 to $\Omega_3=1$. } tabulated in the supplemental material of~\cite{Rougemont:2021qyk}.

\subsection{Remarks on the holographic model construction}
\label{sec:embedding}
Actions of the form~\eqref{eq:action} can arise within extensions of the top-down constructed original AdS/CFT correspondence between $\mathcal{N}=4$ Super-Yang-Mills (SYM) theory and type IIB Supergravity (SUGRA). 
For example, within type IIB superstring theory we may consider adding extra Dirichlet-branes (D-branes) to the standard AdS/CFT-construction involving a stack of $N$ coincident D3 branes. While we will not give an explicit string theory embedding for~\eqref{eq:action}, this is the way we approach its holographic interpretation. 
We consider the gauge theory dual to our holographic model~\eqref{eq:action} as a deformed version of $\mathcal{N}=4$ SYM theory. It is deformed in that it is allowed to have a Chern-Simons coupling, $\alpha$, breaking supersymmetry, and in that it contains an extra symmetry, namely the $U(1)_V$-symmetry that is manifesting in the presence of an extra gauge field $V^\mu$. Recall that $\mathcal{N}=4$ SYM theory does already contain an anomalous $U(1)_A$-symmetry (and associated gauge field $A^\mu$), which is a subgroup of the R-symmetry (the symmetry which rotates supercharges into each other).  

Choosing values for the Chern-Simons coupling $\alpha$ and the gravitational coupling constant $\kappa$ defines the dual field theory. 
In this work, we choose the Chern-Simons coupling such that our field theory has the same chiral anomaly as QCD, namely by $\alpha_{QCD}=6/19$. 
In order to reduce our holographic model to $\mathcal{N}=4$ SYM theory, one would choose the vector gauge field to vanish, $V^\mu\equiv 0$, along with choosing the supersymmetric value of the Chern-Simons coupling, $\alpha=1/\sqrt{3}$.  

While the matching of the anomaly coefficient is straight forward, the matching of the gravitational coupling is not as well constrained. In this work we utilize the AdS/CFT dictionary to give the value of $\kappa$ in terms of the number of colors $N_c$ and the AdS radius. In the most familiar form of the correspondence built as a stack of $N_c$ coincident $D_3$ branes in type IIB string theory one finds
\begin{equation}\label{eq:holo_dict_kappa}
    \kappa^2=\frac{4 \pi ^2 L^3}{N_c^2}\,.
\end{equation}
Unfortunately, as discussed above, it is unclear if the relation given in eq.~(\ref{eq:holo_dict_kappa}) applies to the action of our holographic model~\eqref{eq:action} as it is a deformed version of $\mathcal{N}=4$ SYM theory. Nevertheless, we will use eq.~(\ref{eq:holo_dict_kappa}) as our definition of $\kappa$ in our deformed version of $\mathcal{N}=4$ SYM theory.
A separate method of fixing the value of the gravitational coupling is discussed in~\cite{Ghosh:2021naw}. Their method relies on the fact that the late time geometries in their calculations will be a static black brane. Then, one can consider matching thermodynamic data~\cite{Ghosh:2021naw}. In particular the entropy of the black brane is matched to three-fourths of the Stefan-Boltzmann value for three flavor QCD\footnote{
It has been shown repeatedly that thermodynamic quantities as calculated at strong coupling via the gauge/gravity duality are relatively unaffected by the strong coupling, taking on approximately $75\%$ of their weak coupling values, see for instance~\cite{Gubser:1996de}.}. 

The method used in~\cite{Ghosh:2021naw} poses several problems in our work, perhaps the most challenging of which is that the solutions we obtain will never asymptote to a static black brane configuration. Instead, as stated above, we will assume that the relation in eq.~(\ref{eq:holo_dict_kappa}) continues to hold. We are then forced to fix the number of colors, for which we set $N_c=3$. Although not infinite, lattice studies have shown that $N_c=3$ produces thermodynamic results not so different ($\sim 20\%$) from $N_c=\infty$~\cite{Teper:1998te,Meyer:2003wx,Panero:2012qx}. In addition, we must fix the value of the AdS radius $L$. 
We then make the arbitrary choice of $L=1\,\unit{fm}$. Having fixed the AdS radius and the number of colors we are now in a position to match to QCD data. To do so, we perform one run of our code at arbitrary parameters and obtain the dynamical temperature $T=\epsilon^{1/4}$ in units of ${\rm GeV}$. We then use a scaling symmetry of our field theory, which acts on physical quantities as 
\begin{subequations}
\begin{align}
    \tilde{x}^i &= \frac{x^i}{\lambda}, \quad \tilde{B}=\lambda^2 B, \quad \tilde{\mu}_{(5)}=\lambda \mu_{(5)} \, ,\\
     \vev{\tilde{T}_{\mu\nu}}&=\lambda^4 \vev{T_{\mu\nu}},\,\, \vev{\tilde{J}_{(5)}^i}= \lambda^3\vev{J_{(5)}^i}, \,\, \vev{\tilde{J}^i}= \lambda^3\vev{J^i} \,,
\end{align}
\end{subequations}
to scale the data of our run such that at $\tau=7\, \unit{fm}$ our temperature is $T\approx 90 \,\unit{MeV}$ or, $T(7 \,\unit{fm}/\lambda)\lambda = 90 \,\unit{MeV}$. For the scaling run we used, the value of the rescaling we find is $\lambda=0.768$. Once we have fixed the scaling, we obtain starting values of our initial data such that it yields models of collisions corresponding to RHIC BES beam energies of $\sqrt{s}=( 19\,\unit{GeV},27\,\unit{GeV}, 39\,\unit{GeV}, 64\,\unit{GeV}, 200\,\unit{GeV})$. We also include $\sqrt{s}=2.74\,\unit{TeV}$ in some of our analyses to show the distinction between RHIC and LHC energies in our model. 
\section{Results: CME current \& pressure anisotropies}
\label{sec:results}
In this section we present and analyze the results for the time evolution of the energy density and pressures as well as the CME current.

\subsection{AdS radial dependence of the CME current }
We begin by showing that the time evolution of the energy density, pressures, and CME current is largely independent from the choices of profiles in the AdS radial direction at the initial time. In the dual field theory, the choice of the bulk profile on the initial time slice corresponds to choosing the magnitude of spatial anisotropies as well as their initial time-derivative. The choices of this initial data, given in eq.~(\ref{eq:init_Data}) at fixed axial charge, magnetic field, initial time, and initial energy density, are shown in the top panel of figure~\ref{fig:initial_data}. The results of our time evolution for the energy-momentum tensor and the vector current are displayed in the middle and bottom panels of figure~\ref{fig:initial_data}, respectively. 
In the middle panel of figure~\ref{fig:initial_data} we see that the difference in initial profile leads to variations of the pressures and the energy density where we have highlighted a ``characteristic'' curve with a thicker line with the remaining initial conditions represented as thinner lines. However we see that the variations die off over time. In the pressures the variations persist until approximately $\tau T=2.2$. While for the energy density the variations persist until approximately $\tau T=3$, beyond which it is seen that all the curves evolve with approximately the same behavior, despite being at different values of energy density for any given set of initial conditions.

However, one can see that the different initial data in general does not have a very strong effect on the time evolution of the energy-momentum tensor. This is especially pronounced in the bottom panel of figure~\ref{fig:initial_data}, where we have displayed the vector current (CME current) for each of the different initial conditions. The CME response is clearly robust against different choices of the initial spacetime anisotropy. Here, again, we have shown a ``characteristic'' curve with a thicker line and this curve corresponds to the same characteristic curve as displayed in the upper two panels of figure~\ref{fig:initial_data}.

\subsection{Energy dependence of the CME}  
Having shown that our results do not depend strongly on the choice of initial data for the spacetime anisotropy, we fix a particular form of the initial data in eq.~(\ref{eq:init_data}) with 
\begin{subequations}
\begin{align}
    \Omega_i&=\gamma_i=\beta_3=\beta_4=\beta_5=0 \, , \\
    \beta_0&=1/10,\quad \beta_1=2/10 \quad \beta_2=-1/2 \,.
\end{align}
\end{subequations}
 This choice of initial parameters for the initial spacetime anisotropy profile corresponds to the thick lines displayed in figure~\ref{fig:initial_data}. We then use the matching procedure outlined in section~\ref{sec:embedding} to present our results in physical units. 
 
In the ensuing sequence of plots and titled paragraphs we display the results of the analysis for various energies relevant to the RHIC beam energy scan. We provide this analysis in several stages. We begin by holding all initial input values fixed at the initial time while we vary the initial energy density at the QCD-matched value for the anomaly coefficient, {i.e.}, $\alpha=\alpha_{\text{QCD}}=6/19$ (case I). 
We then repeat this analysis for a second value of the anomaly coefficient, the supersymmetric value for which $\alpha=\alpha_{\text{SUSY}}=1/\sqrt{3}$ (case II). 
After this we return to the physically relevant QCD value for the anomaly coefficient while we vary the initial energy and scale the initial axial charge density accordingly (case III). We then study the dependence on the initial magnetic field value by repeating our analysis with an initial magnetic field of half the size (case IV). We then transition and consider the magnetic field on the initial time slice to depend on the initial energy density while holding the initial axial charge fixed (case V). Finally we push our model to its logical conclusion by varying the initial energy while scaling the initial axial charge density and initial peak magnetic field strength accordingly (case VI). \\

\begin{figure}[htbp]
\centering
    \includegraphics[width=0.47\textwidth]{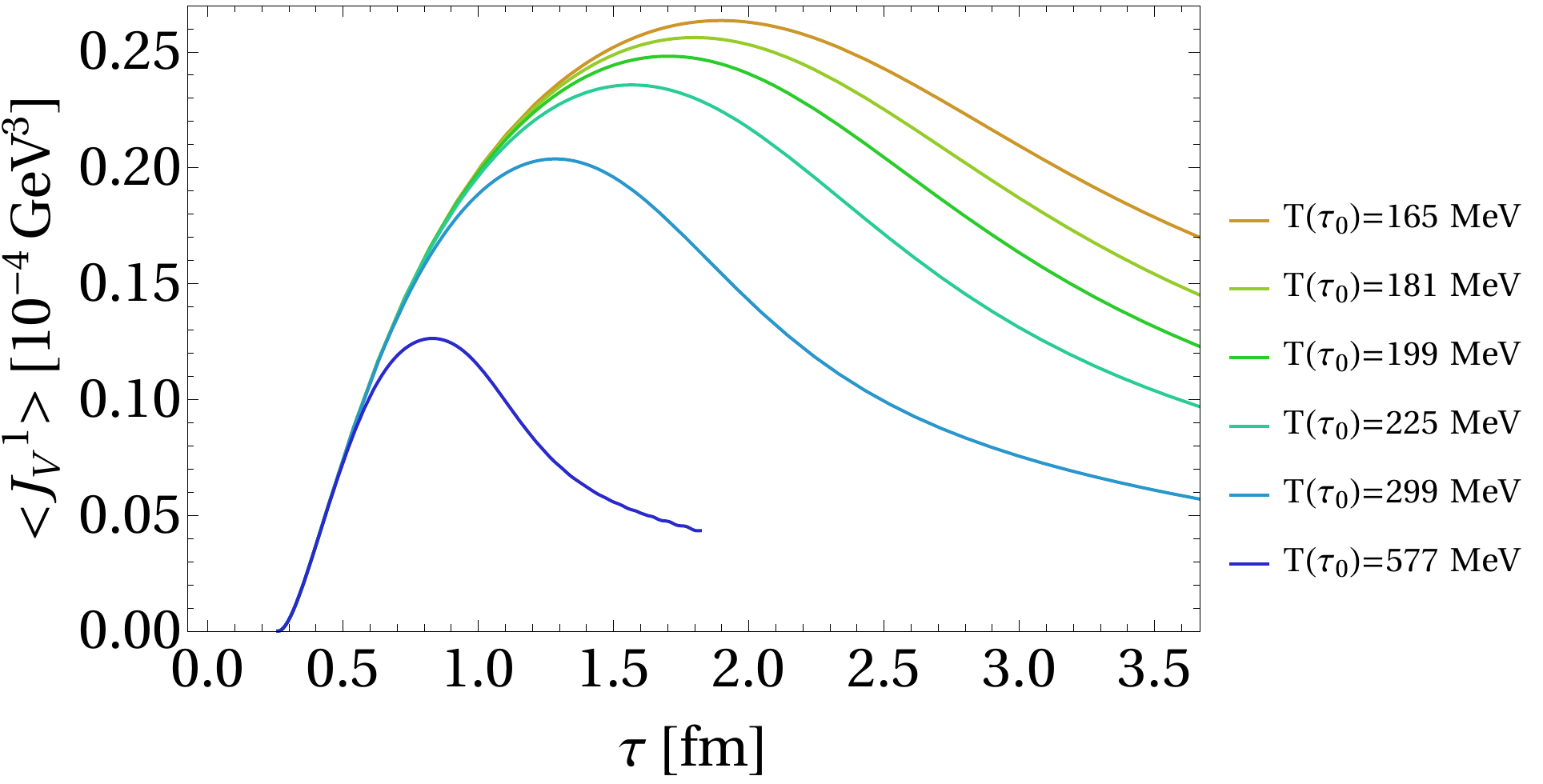} 
   \caption{
   {\it CME current -- fixed initial axial charge and magnetic field strength (case I):} The vector current $J_V^1$ along the magnetic field is displayed for six different beam energies (initial energy densities) for a fixed initial axial charge density of $\vev{J^0_{(5)}}= 0.00032 \,\unit{GeV}^3$, and the initial magnetic field is fixed to be $e B\approx m_\pi^2$. 
   \label{fig:vec_current}}
\end{figure}

\noindent \textbf{Case I - Fixed initial axial charge and magnetic field strength: } In figure~\ref{fig:vec_current} we display the results of our simulations for a fixed initial value of the magnetic field of $e B\approx m_\pi^2$ and axial charge density\footnote{In Case III we will see that this is actually an order of magnitude smaller than estimates of the axial charge density we use in later plots. However, here this value is chosen simply to illustrate the basic behavior of our system. Using a value of the same order of magnitude as in Case III-VI does not change the essential physics, it only scales the magnitude of the CME response.} $\vev{J^{0}_{(5)}} =0.00032 \,\unit{GeV}^3$, 
at initial energy densities corresponding to values of collision energy and temperature as displayed in table~\ref{tab:energy},
\begin{table}[h]
\caption{Starting values of energy density and temperature used in our numerical simulations. \label{tab:energy}}
\begin{ruledtabular}
  \begin{tabular}{lllllll}
        $\sqrt{s}\hspace{0.1cm}[\unit{GeV}]$  & $19$ & $27$& $39$& $64$ & $200$ & $2750$\\
      $T\hspace{0.1cm} [\unit{MeV}]$ & $165 $ & $181 $ & $199$& $225$ & $299$ & $577$  
    \end{tabular}
\end{ruledtabular}
\end{table}
Qualitatively similar for all initial energy densities, the CME current quickly rises to a peak value before decreasing with time. Remarkably, the peak value obtained for the highest initial energy we used (blue curve) is smaller than that of the lowest initial energy run (orange curve). In addition, the highest initial energy run has the sharpest fall-off for the CME current. Hence, we conclude that for this set of initial data the CME current survives longer at lower beam energies. At the end of this section, we will quantify these statements by introducing a measure of the total amount of CME current that has flowed per area. This measure is referred to as {\it charge accumulation} and was shown already in figure~\ref{fig:charge_totals}. 
%

\begin{figure}[htbp]
    \centering
    \includegraphics[width=0.45\textwidth]{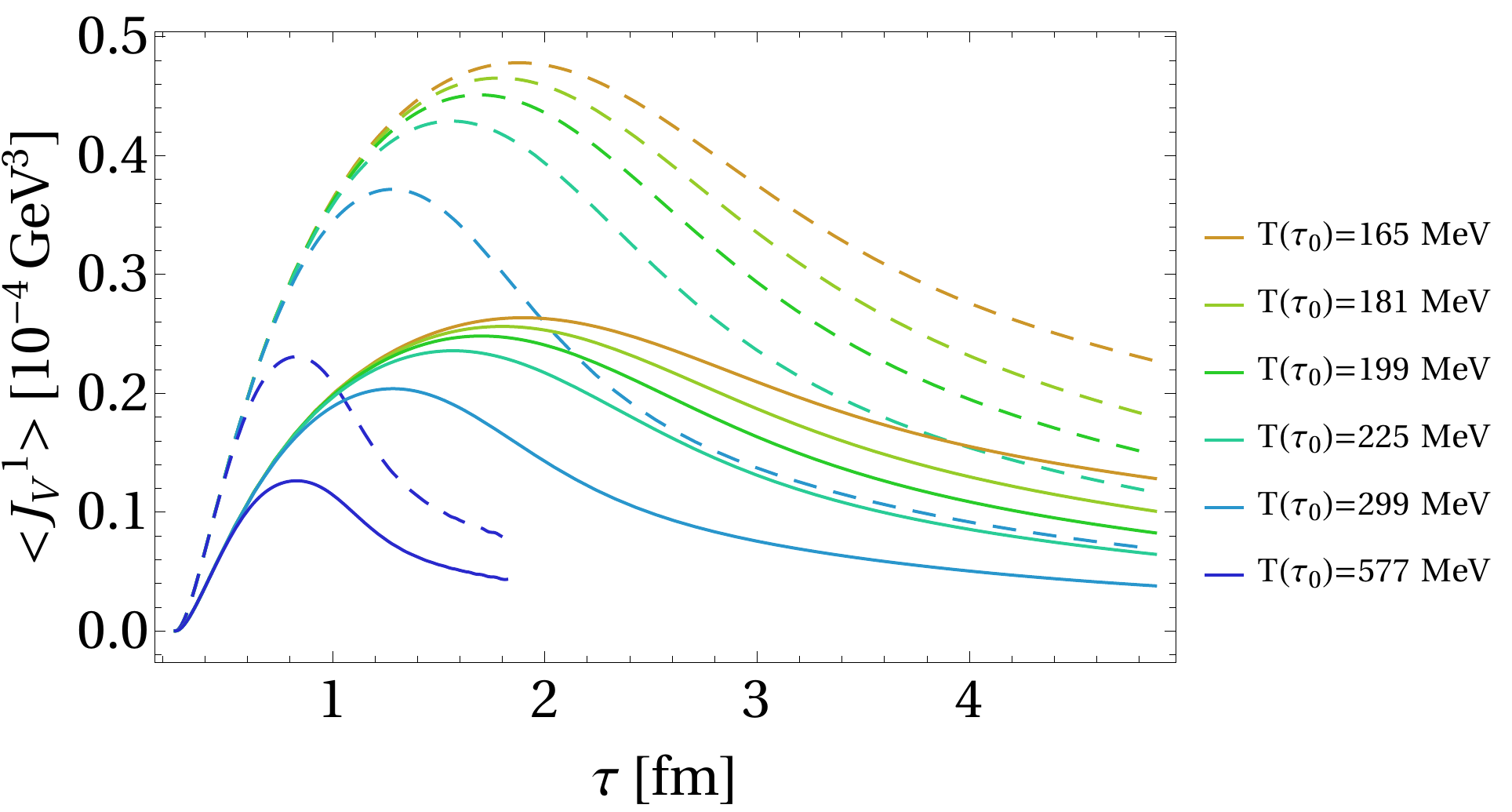}
    \caption{{\it CME current -- anomaly coefficient dependence (case II):} The vector current $J_V^1$ along the magnetic field is displayed for four different beam energies (initial energy densities) for a fixed initial axial charge density. The solid curves are generated with $\alpha_{\text{QCD}}\approx 0.319$ while the dashed curves are generated at the supersymmetric value of $\alpha=1/\sqrt{3}$. In both cases the initial axial charge density is $\vev{J^0_{(5)}}= 0.00032 \,\unit{GeV}^3$, while the initial magnetic field is fixed to $e B\approx m_\pi^2$.
    \label{fig:SUSY_QCD}}
\end{figure}

\noindent \textbf{Case II - Anomaly coefficient dependence:} It is interesting to ask how the value of the anomaly coefficient affects the response of the CME current. In figure~\ref{fig:SUSY_QCD} we address this question by comparing the evolutions for $\alpha_{\text{QCD}}=6/19\approx 0.316$ and the formerly supersymmetric value $\alpha_{\text{SUSY}}=1/\sqrt{3}\approx 0.577$. In figure~\ref{fig:SUSY_QCD}, the larger supersymmetric value of the anomaly coefficient (dashed curves)   
leads to an enhancement of the peak signal (and of the whole curve at all times) by an approximate factor of two. This is consistent with the ratio $\alpha_{\text{QCD}}/\alpha_{\text{SUSY}}\approx 0.54$ as expected from eq.~(\ref{eq:axialcurrent}) since $C \propto \alpha$. 

\noindent \textbf{Case III - Varying initial energy and axial charge densities at fixed initial magnetic field: }
The peak value of the axial charge density generated during a collision changes as a function of the beam energy. Since our model contains a time-dependent charge density it is then interesting to consider appropriately changing the initial axial charge density along with the initial energy. As shown in~\cite{Sun:2018idn}(and the references therein) the axial charge density is related to fluctuations of color electric and magnetic fields. We collect their arguments here and display the resulting axial charge density approximated as follows~\cite{Sun:2018idn}: 
\begin{equation}
    2\vev{J^{0}_{(5)}}\tau_{th} A_{\text{overlap}}=\frac{2\tau_{th}^2\pi\rho_{\text{tube}}^2Q_s^4\sqrt{N_{\rm tube}}}{16\pi^2}\approx 135\,, \label{eq:estimateAxialCharge}
\end{equation}
\begin{figure}[H]
\centering
    \includegraphics[width=0.45\textwidth]{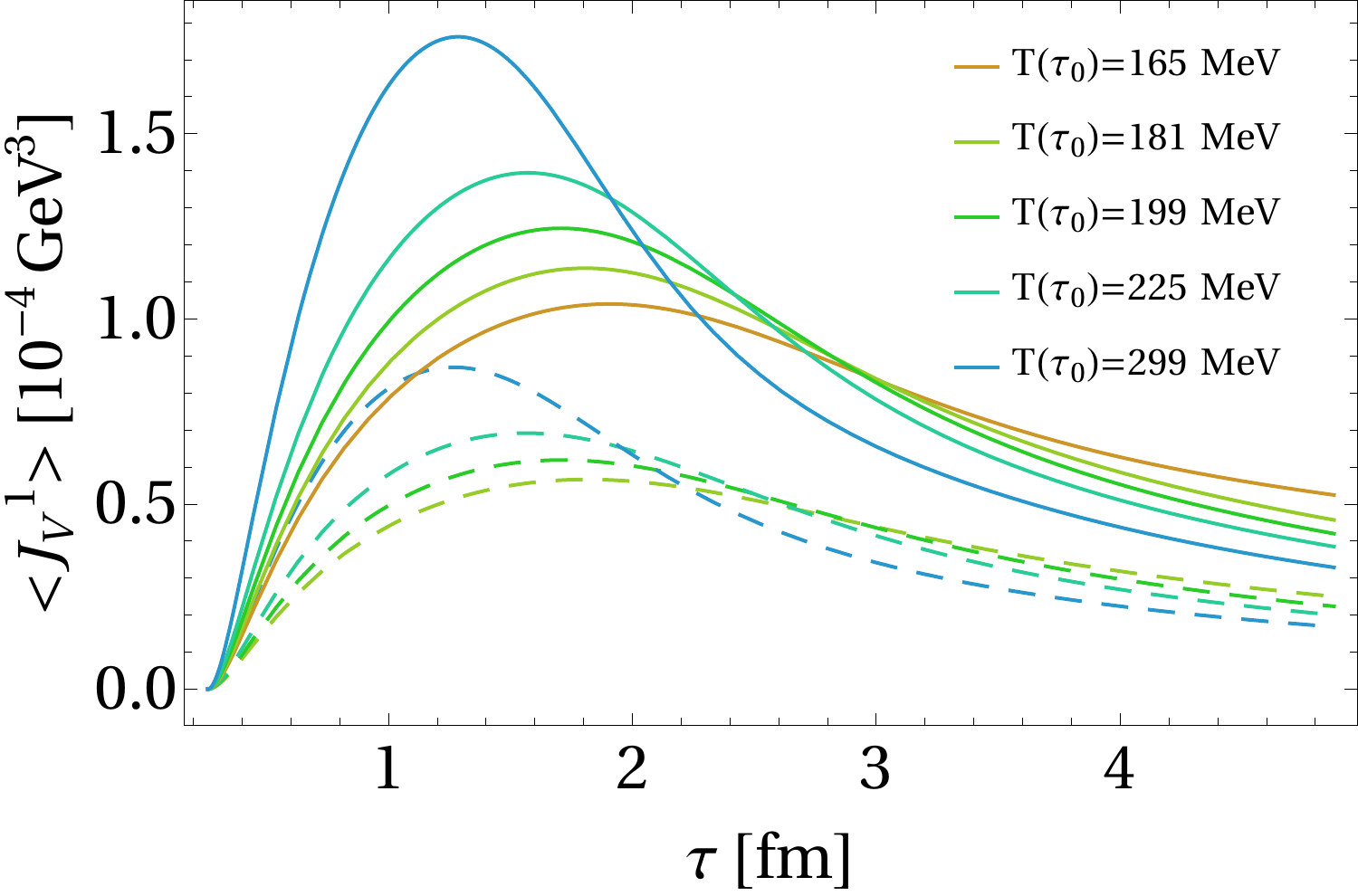}
  \caption{{\it CME current -- varying initial axial charge densities (cases III and IV):} The vector current $J_V^1$ along the magnetic field is displayed for four different beam energies (four different initial energy densities) for an initial axial charge density which scales with the initial energy density as displayed in table~\ref{tab:axial}. The solid curves (case III) are generated with $e B\approx m_\pi^2$ and dashed curves (case IV) are generated with $e B\approx m_\pi^2/2$. 
  \label{fig:vec_current_charge_scaling}}
\end{figure}
where $\rho_{\text{tube}}$ and $N_{\text{tube}}$ are the radius and number of glasma flux tubes, with the latter approximately equal to the number of binary collisions between nuclei $N_{\text{tube}}\approx N_{\text{coll}}$, and $\rho_{\text{tube}}\approx 1 \,\unit{fm}$ at $200\,\unit{GeV}$. 
The other parameters are the thermalization time $\tau_{th}$ of the quark gluon plasma (QGP) which is taken to be $\tau_{th}=0.6\,\unit{fm}$ along with the saturation scale $Q_s^2\approx 1.25\, \unit{GeV}^2$ for gluons at collision energy of $200\,\unit{GeV}$. Finally the number of collisions can be estimated~\cite{Sun:2018idn} to be $N_{\text{coll}}=82.73$ for $Zr+Zr$ and $Ru+Ru$ for impact parameter $b=7\,\unit{fm}$. We can then determine the axial charge density by considering a rough estimate for the area of overlap $A_{\text{overlap}}$ as a circle of radius $10\,\unit{fm}$ to give  $\vev{J_{(5)}^0}= 0.0027\, \unit{GeV}^3$. 

Given this estimate of the value of the axial charge at $200\,\unit{GeV}$ we must now produce an estimate for its value for all energies we consider. We arrive at these values by using that the axial charge density scales as $Q_s^2$ (when assuming that $\rho^2_{\rm tube} \sim 1/Q_s^2$) and $Q_s^2 \sim (1/x)^{1/3} \sim s^{1/3}$, and hence we arrive at the values displayed in table~\ref{tab:axial}. 
\begin{table}[h]
\caption{Estimates of axial charge density at various beam energies (initial energy densities) used in the numerical evolution. \label{tab:axial}}
\begin{ruledtabular}
  \begin{tabular}{lllllll}
        $\sqrt{s}\hspace{0.1cm} [\unit{GeV}]$  & $19$ & $27$& $39$& $64$ & $200$ & $2750$\\
        $\vev{J^0_{(5)}}\hspace{0.1cm} [10^{-3}\unit{GeV^3}]$ & $1.2 $ & $1.4 $ & $1.6$& $1.9$ & $2.7$ & $6.6$  
    \end{tabular}
\end{ruledtabular}
\end{table}
There are a few caveats associated with the values given in table~\ref{tab:axial}. First, each of these values we calculated assuming a thermalization time of $\tau_{th}=0.6 \,\unit{fm}$. 
Our simulation begins before this thermalization time occurs. To be proper, one should use the expected time evolution of the axial charge in our setup (see eq.~(\ref{eq:dual_currents_A})) to propagate these values back to the initial time slice. Doing so increases the values reported in table~\ref{tab:axial} by a factor of three. Given that each of the values would be scaled by an overall factor of three we expect only a scaling of the resulting CME current by an approximate factor of three for all curves. 
In principle, we could run these simulations. However, our holographic method encounters a systematic problem for such large values of the axial charge density at the initial time, as this set of initial data leads to situations where it is not possible to form an apparent horizon.\footnote{This fact may indicate that this system needs all the energy to support the required charge density, similar to an extremal Reissner-Nordstr\"om black hole, which carries charge but has no thermal energy (although it is an equilibrium state which has an event horizon). Increasing the initial axial charge density even further may lead to unphysical situations where the supplied energy density is insufficient to support the demanded axial charge density. However, there may also be interesting dynamical solutions possible in this ever-expanding setup, which have no equilibrium analogs. We leave this for future investigations. } 
Hence, we choose not to do this and use the values reported in table~\ref{tab:axial} as the data on the initial time slice. 
What is important for interpreting our results is the relative scaling with energy, not the overall scaling as a result of back-propagation. The results of this simulation are displayed in figure~\ref{fig:vec_current_charge_scaling} as solid lines. We find that with the energy-dependent initial axial charge density the higher energy collisions lead to a faster rise and a larger peak, but also a faster decrease of the CME current.  \\


\noindent\textbf{Case IV - Decreased magnetic field strength:} Here we have repeated the analysis of case III with half the initial value of the magnetic field (i.e., $e B(\tau_0)=m_\pi^2/2$). The results of this analysis are displayed as dashed lines in figure~\ref{fig:vec_current_charge_scaling}. This comparison is made to show that simply decreasing the initial value of the magnetic field for all sets of initial data by the same amount has a minimal effect on the resulting CME current, apart from its overall reduction. Given that both the axial charge and magnetic field in the field theory are time-dependent, it is a non-trivial outcome of this test to see that the peaks of the CME current shift slightly as we decrease the initial magnetic field strength. For $e B(\tau_0)=m_\pi^2/2$ the CME current peaks later for $\sqrt{s}=(27,39)\,\unit{GeV}$ while they peak sooner for $\sqrt{s}=(64,200)\,\unit{GeV}$ as compared to the case of $e B(\tau_0)=m_\pi^2$. \\

\noindent\textbf{Case V - Varying initial energy density and magnetic field strength at fixed axial charge density:} While it is still unclear what the appropriate magnitude, spatial extent and time evolution of the magnetic field generated during a heavy ion collision is, it is well agreed upon that its magnitude changes as a function of the beam energy. Since our model contains a time-dependent magnetic field, determined by symmetries and equations of motion alone, it is then interesting to consider appropriately changing the initial magnitude of the magnetic field along with the beam energy. In this case we work again at a fixed value of the axial charge density on the initial time slice as a function of the energy density. We do so in order to isolate the effect of an initial energy-dependent peak magnetic field strength on the CME current.

\begin{figure}[H]
\centering
    \includegraphics[width=0.45\textwidth]{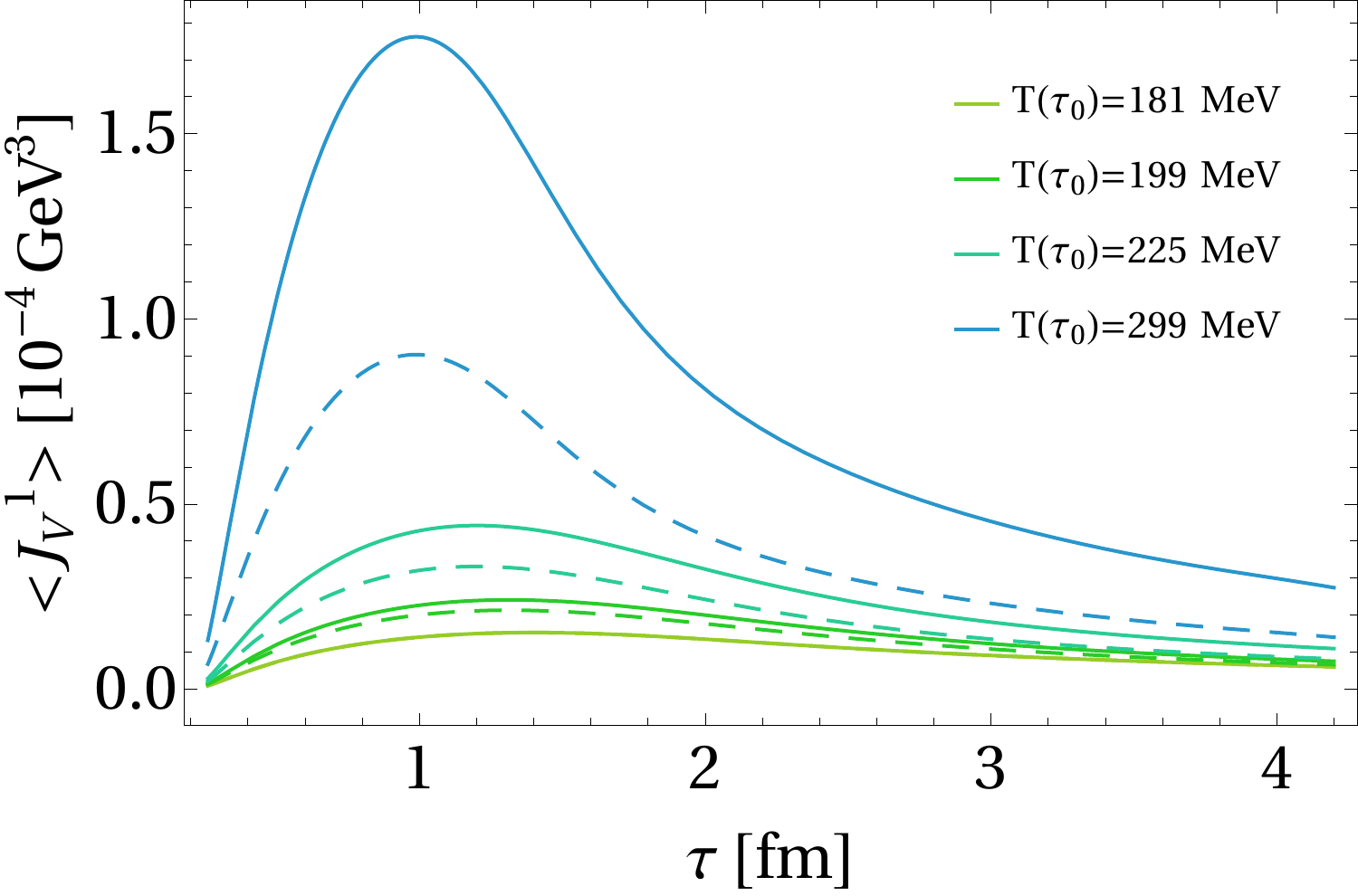}
  \caption{{\it CME current (cases V and VI):} The vector current $J_V^1$ along the magnetic field is displayed. The dashed lines (case V) indicate the time evolution for four different beam energies (initial energy densities) for an initial axial charge density independent of the initial energy density and magnetic field which scales with the initial energy as shown in table~\ref{tab:Mag_field_Strength}. Solid lines (case VI) indicate running for four different energies with an initial axial charge density and magnetic field which scales with the initial energy as shown in table~\ref{tab:Mag_field_Strength} and table~\ref{tab:axial}. 
  \label{fig:vec_current_charge_mag_scaling}}
\end{figure}

\begin{figure*}[t]
\centering
    \includegraphics[width=0.95\textwidth]{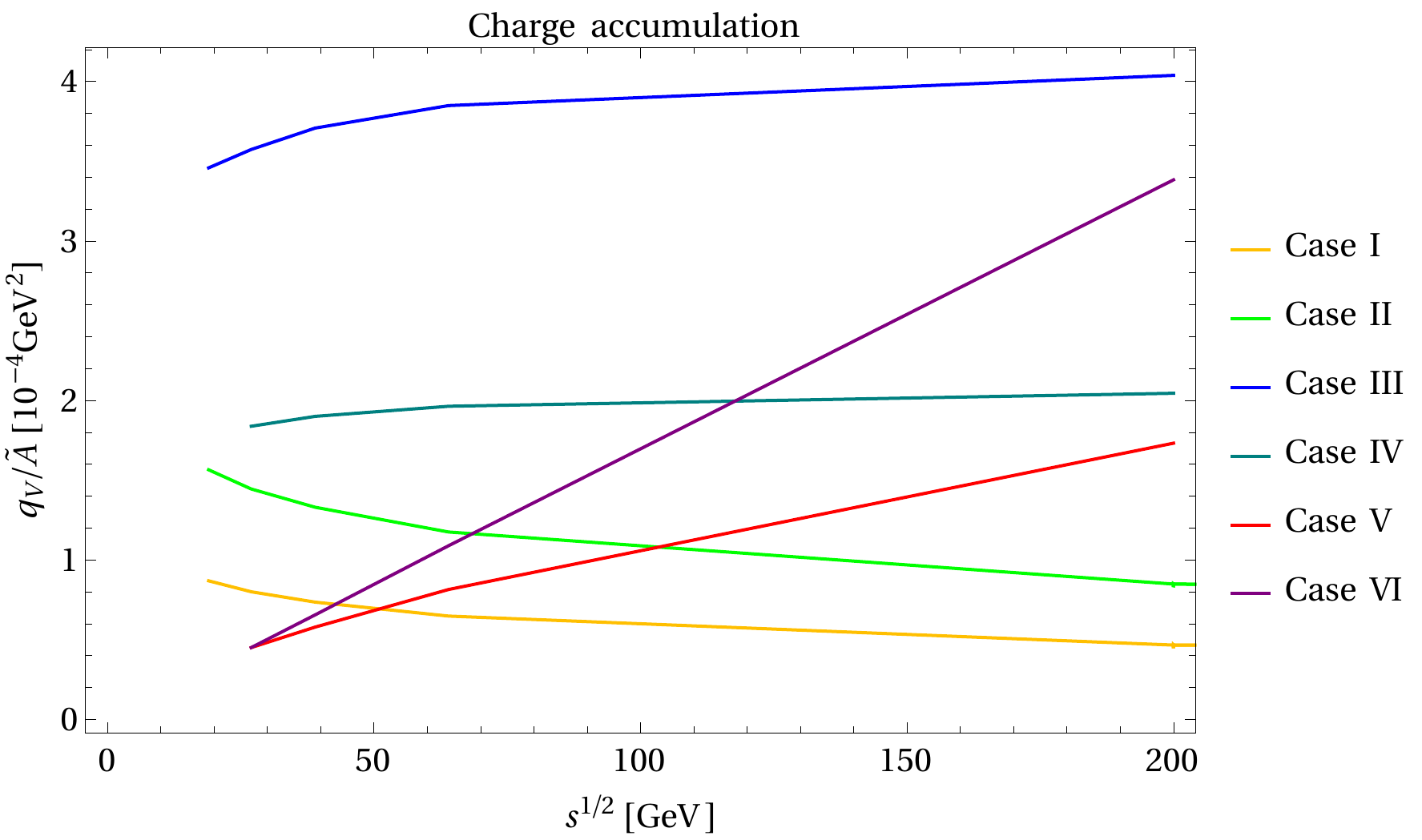}
  \caption{
  {\it Charge accumulation from time-integrated CME current:} The total amount of charge per area which has flowed during the simulations considered throughout this work (see eq.~(\ref{eq:time_int_CME})). The plot legend labels the case in which the total charge was computed corresponding to the titled paragraphs in section~\ref{sec:results}. The cases differ by either holding fixed, or varying, the initial value of the magnetic field and the axial charge density, ($B_1(\tau_0),\vev{J_{(5)}^0(\tau_0)}$), as a function of the initial energy density at the initial time, $\tau_0$. Case I: both $\vev{J_{(5)}^0(\tau_0)}$ and $B_1(\tau_0)$ are constant as a function of initial energy. Case II: both $\vev{J_{(5)}^0(\tau_0)}$ and $B_1(\tau_0)$ are constant as a function of initial energy while the Chern-Simons coupling $\alpha$ is taken at the supersymmetric value. Case III: $B_1(\tau_0)$ is held fixed while $\vev{J_{(5)}^0(\tau_0)}$ varies as a function of initial energy density. Case IV: case III is repeated with $B_1(\tau_0)$ taking  half the value of case III. Case V: $\vev{J_{(5)}^0(\tau_0)}$ is held fixed while $B_1(\tau_0)$ varies as a function of initial energy density. Case VI: both $\vev{J_{(5)}^0(\tau_0)}$ and $B_1(\tau_0)$ vary as a function of the initial energy density. 
  \label{fig:charge_totals}}
\end{figure*}

A simple estimate of the peak magnetic field strength at the center of the collision is given in~\cite{McLerran:2013hla} as 
\begin{equation}\label{eq:Magnetic_field_energy_Dependence}
   eB=\frac{1}{2}\frac{\gamma_{*}}{\gamma}\left(\frac{Q_s}{Q_s^{*}}\right)^2 \left(e B_{*}\right) \, ,
\end{equation}
where $B$ is the magnitude of the magnetic field, which we wish to compute at energies lower than a given high energy scale. Quantities at that high energy scale are starred. Here $\gamma$ is the Lorentz factor of the associated collision energy and $Q_s$ is the gluon saturation scale. In this work, in order to compare as closely as possible to previous work~\cite{Ghosh:2021naw}, we choose to fix the high energy magnitude of the magnetic field to $eB_*=10m_\pi^2$ at $1\ui{TeV}$. Application of this formula then provides us with the estimates for the magnetic field strength shown in table~\ref{tab:Mag_field_Strength}.
\begin{table}[h]
\caption{Estimates of peak magnetic field strengths at various beam energies (initial energy densities). \label{tab:Mag_field_Strength}
}
\begin{ruledtabular}
  \begin{tabular}{llllll}
        $\sqrt{s} [\unit{GeV}]$  & $19$ & $27$& $39$& $64$ & $200$ \\
        $eB$ & $0.095m_\pi^2$ & $0.135 m_\pi^2$ & $0.195 m_\pi^2$ & $0.32 m_\pi^2$ & $1 m_\pi^2$  \\ 
    \end{tabular}
\end{ruledtabular}
\end{table}
The results of our analysis are displayed as dashed lines in figure~\ref{fig:vec_current_charge_mag_scaling}. We see that by holding the charge density fixed, independent of the initial energy while varying the initial magnetic field as a function of the energy, the CME current response is similar to that displayed in figure~\ref{fig:vec_current_charge_scaling}. Large initial energies lead to larger peak values of the CME current with sharper fall-offs over time. \\

\noindent\textbf{Case VI - Varying initial energy density, magnetic field strength and axial charge density:} As a final case we consider varying both, the magnetic field and the axial charge density, with the initial energy density (modeling different beam energies). The results of this analysis are displayed in figure~\ref{fig:vec_current_charge_mag_scaling} as solid lines. We can see that the peak value is the largest for the highest energy ($200\,\unit{GeV}$ collisions). This peak value is quickly damped away, but we show below that the large peak will manifest itself in a large total charge, accumulated during the evolution, and contributes to the main conclusion of this work (see figure~\ref{fig:charge_totals}). In contrast to case V, shown as dashed lines, the overall magnitude of the signal is larger. \\

\noindent \textbf{Charge accumulation:} Although the previous results already indicate how the vector current varies throughout its time evolution depending on the initial conditions, it will be useful to assign some measure of how much current is produced for a given case. 
This will provide an estimate of the total amount of CME-produced charge separation that would be detectable.
A simple, yet effective measure is the amount of charge that flows through a surface of a given area throughout the duration of the simulation, 
\begin{equation}
    q_V=\int \exd t\int \vev{\vec{J}}\cdot\exd\vec{A} \, .
\end{equation}
Given that the current flows along the $x_1$ direction we can take $\exd \vec{A}=\hat{i}\exd x_2\exd x_3$ from which we find 
\begin{equation}
      q_V=\int \exd t \exd x_2 \exd x_3 \vev{J^1}= \int\exd x_2\exd \xi \int\exd\tau \tau  \vev{J^1}\,.
\end{equation}
The bounds on the integrals for $x_2$ and $\xi$ are our choice, so we will choose to compute this per unit value of the area $\tilde{A}$ defined as $\tilde{A}=\int \exd x_2\exd\xi$, and report the {\it charge accumulation} defined as 
\begin{equation}\label{eq:time_int_CME}
    q_V/\tilde{A}=\int\limits_{\tau_i}^{\tau_f}\exd\tau \tau  \vev{J^1},
\end{equation}
considering it a measure of the charge which flows through a unit area throughout the duration of our simulation. 
The values of $\tau_i=0.260\, \unit{fm}$ and $\tau_f=4.88 \, \unit{fm}$, the initial and final time, respectively, correspond to the duration of each simulation (in cases I through VI). 
The results of this calculation are displayed in figure~\ref{fig:charge_totals}.

The results show that only two of the cases we consider in this work produce a larger amount of charge transport as we decrease the energy of the collisions (cases I and II). These curves are displayed as the orange and bright green lines in figure~\ref{fig:charge_totals}. 
They correspond to the case of working with values of the magnetic field and axial charge density at the initial time, which do not depend on the initial energy density. The difference in the two curves is due to the choice of the Chern-Simons coupling, the orange curve corresponding to the case with $\alpha=\alpha_{QCD}$ (case I), the bright green corresponding to $\alpha=\alpha_{SUSY}$ (case II). Using the supersymmetric value increases the amount of charge accumulated, but the trend still displays a decreasing slope as a function of the initial energy density. 

The blue line in figure~\ref{fig:charge_totals} corresponds to case III, where the initial magnetic field is independent of the choice of initial energy density, but the value of the initial axial charge density scales with the intial energy density according to the model shown in eq.~\eqref{eq:estimateAxial}~\cite{Sun:2018idn}. In this case our model predicts that an increased amount of charge will pass through a unit surface in the field theory during the evolution as we increase the initial energy density. 
The forest green line corresponds to case IV, where we have repeated case III with only the magnetic field reduced by a factor of two. Here we clearly see the influence of the choice of the initial value of the magnetic field if it is kept independent of the initial energy density. An increased value of the magnetic field leads to an increase in current flow as expected already from the near-equilibrium current eq.~\eqref{eq:axialcurrent}. 

The red curve in figure~\ref{fig:charge_totals} corresponds to case V, working with an initial axial charge density, independent of the initial energy while allowing the magnetic field on the initial slice to vary as given in eq.~(\ref{eq:Magnetic_field_energy_Dependence}). Here, we again see an increasing amount of total charge has passed through a surface during our simulation as we increase the energy of the collision. 

The final curve (purple) which we display in figure~\ref{fig:charge_totals} corresponds to case VI, allowing both the magnetic field and axial charge density to depend on the initial energy density (beam energy). Here, we see the largest positive slope, indicating a significant increase in the accumulated CME charge as a function of the initial energy density.


\textbf{Comparison to previous work:} 
Just like the present work, earlier works have also considered fully backreacted setups, i.e., setups in which the bulk matter sources the Einstein equations. Among them are time-dependent holographic models including the  works~\cite{Cartwright:2020qov,Ghosh:2021naw}. The model used in~\cite{Cartwright:2020qov} is similar to the one used in this work, however, in~\cite{Cartwright:2020qov} the model did not contain a vector gauge field, $V_{\mu}=0$. As a result, the model used in~\cite{Cartwright:2020qov} is a consistent truncation of type IIB string theory, provided the supersymmetric value of the Chern-Simons coupling is chosen. It further differs in construction by choice of boundary conditions on the metric, where the author in~\cite{Cartwright:2020qov} enforced that,
\begin{equation}\label{eq:flat_BC}
\lim_{r\rightarrow\infty}   \frac{L^2}{r^2} g_{\mu\nu}=\text{diag}(-1,1,1,1) \, ,
\end{equation} 
implying that the geometry of the boundary field theory is flat Minkowski spacetime. With this choice of boundary metric and the choice that field theory quantities should depended only on time, the author's chosen initial data generates initially anisotropic plasmas with trivial flow (i.e., no expansion). Furthermore the author's choice of axial gauge field ansatz leads to aligned, time-independent, external electric and magnetic fields in the dual field theory. The consequence of this choice is an indefinite production of axial charges. Due to these differences, in particular since this model does not account for any vector current, it is difficult to compare the results of~\cite{Cartwright:2020qov} with those of the present work. 

The model most similar to the one used in the present work is that of~\cite{Ghosh:2021naw}, where the authors furthered the work of~\cite{Cartwright:2020qov} to include the missing $U_V(1)$ gauge field and used the same model as presented in eq.~(\ref{eq:action}) to study the evolution of the vector, rather than axial, current response in an isotropizing SYM plasma. To achieve this particular set of solutions, the authors use the same boundary conditions on the metric as the author of~\cite{Cartwright:2020qov} (see eq.~(\ref{eq:flat_BC})) leading to a flat boundary spacetime. With this choice of boundary metric and the choice that field theory quantities should depend only on time, the authors choose initial data generating initially isotropic plasmas with trivial flow (i.e., no expansion). The initially isotropic plasma, with non-zero axial chemical potential subjected to an external magnetic field, then relaxes to a final anisotropic configuration with non-zero vector current response. The authors provide a parameter space scan over different energies, axial chemical potentials, anomaly coefficients and magnetic field strengths. 

In contrast to our work, the axial charge density and magnetic field in their work are static. The combination of static axial charge density and magnetic field as well as the static behavior of the fluid (no expansion) makes direct comparison of our results impossible. Looking to the results~\cite{Ghosh:2021naw}, one first observes that while their vector current response grows, and eventually saturates, ours peaks and then decreases over time. This can be attributed to the longitudinal expansion, causing a decreasing axial charge density and magnetic field in our setup. Besides this notable difference, all basic qualitative checks behave as expected. As an example, increasing the value of the anomaly coefficient from the QCD-matched value to the SUSY value leads to an increase in the magnitude of the CME response (as seen in figure~\ref{fig:SUSY_QCD}), the same can be seen in figure 4 of their work.

An important conclusion from their model is that the CME is not fast enough at LHC energies~\cite{Ghosh:2021naw}. This conclusion is based on the observation that the time it takes for their CME current to reach its steady state value is larger than the expected lifetime of the magnetic field. Hence, the authors concluded that their simulation predicts no measurable CME current at LHC energies. In contrast to this, while we have a time-dependent magnetic field within our model we do not draw the conclusion of the viability of the CME current based on lifetimes. Rather, we base our conclusion on the amount of current generated throughout the duration of our simulation. We believe this is a better measure, as what is experimentally relevant is the amount of charged particles which will reach the detectors. 

We note, that we do not see the oscillations reported in \cite{Ghosh:2021naw}. The oscillatory behavior of the one point functions in \cite{Ghosh:2021naw} are interpreted as being due to quasinormal modes of the black brane in their geometry. For large values of the axial charge density, relative to the appropriate powers of the energy density and magnetic field strength, the frequencies of these modes approach the real axis in the complex frequency plane, hence these excitations represent long lived waves/oscillations. 

Our current understanding is that by the time in our evolution that one might expect oscillations in the pressures and the vector current response in the non-expanding case, the value of the magnetic field, axial charge density and energy density have already seen an appreciable decrease in their values. When charge densities become small compared to the temperature, the potential barrier near the horizon shrinks such that the quasinormal modes dissipate more energy into the black hole, leading to a larger damping of the corresponding field theory excitations~\cite{Kaminski:2009ce,Janiszewski:2015ura}. Decreasing the magnetic field below the temperature scale has a similar effect~\cite{Janiszewski:2015ura}. As a result the frequencies of the would be long-lived modes have, rather than approached the real axis, retreated deeper into the complex plane. 
Therefore we see no oscillations as the frequencies of these modes are representative of highly damped excitations.

\section{Discussion}
\label{sec:discussion}
In this work we reported on the beam energy dependence of the chiral magnetic effect in an expanding quark gluon plasma (QGP) based on a holographic model, initially far from equilibrium. Of the six combinations of initial data used, we have found four of them to lead to a larger chiral magnetic effect (CME) signal at higher energies. These are cases III, IV, V, VI for which the CME current is shown as a function of time in figures~\ref{fig:vec_current_charge_scaling} and \ref{fig:vec_current_charge_mag_scaling}. That mentioned CME signal is characterized by the total charge transported along the magnetic field direction (see figure~\ref{fig:charge_totals}), representing the amount of charged particles reaching the detectors during a heavy ion collision. In addition, the only scenario in which the accumulated charge per area is larger at smaller energies represents perhaps the most physically inaccurate description, one in which the initial value of the axial charge density and magnetic field strength, present moments after the initial collision, is independent of the energy of the colliding nuclei; these are cases I and II with the CME currents shown in figures~\ref{fig:vec_current} and~\ref{fig:SUSY_QCD}. 

The conclusion that a more significant CME signal is to be found at higher collision energies is of course based on a model, which necessarily includes caveats. Besides the choice of using a holographic plasma, the model relies on the choice of parameters. We find that the parameters with the strongest influence on the results are the value of initial magnetic field and initial axial charge density at lower energies. In our model these are spatially uniform quantities with dependence only on the proper time $\tau$. Clearly, this is not the case in a realistic heavy ion collision, where the magnetic field and axial charge density can have a complicated spatial distribution and evolution. 
As a result, our current setup with its magnetic field decreasing as $\tau^{-1}$ likely does not correctly reflect realistic relative sizes of the magnetic field, axial chemical potential and energy density at all times. 
This is an especially relevant criticism given there still has to be reached a community consensus on both, the true time dependence of the magnetic field as well as the energy/time dependence of the axial charge density. 
Furthermore, electric and magnetic fields are external and not dynamical in our setup.

An additional source of uncertainty in our results lies in the use of a boost-invariant~\footnote{We recall, that the whole system (action and metric solution) is boost invariant along the $x_3$-direction. That is a symmetry. Nothing depends on $\xi$. The time-evolution of fluid velocity, energy density and pressures is {\emph{not}} Bjorken flow at all times. It is approximately Bjorken flow at late times only.} ansatz.
While this may be expected to be an appropriate approximation at very high energy, it is not expected to be so at the lowest RHIC energies. Furthermore it has been shown previously that Chern-Simons fluctuations are suppressed in the Bjorken rapidity invariant regime~\cite{Kharzeev:2001ev}. There, the authors found that there must be a deviation from the rapidity-invariant regime for Chern-Simons fluctuations to occur. With these deviations expected to be large during low energy collisions there may be a sizable increase in the initial axial charge density present during low energy collisions. 
In addition, at lower beam energies the collision's trajectory through the phase diagram may be closer to the QCD critical point, where the Chern-Simons fluctuations generating the axial charge were recently shown to be enhanced \cite{Ikeda:2020agk}. In our present work these two potential sources of additional initial axial charge were not considered and taken together may provide enough initial axial charge density to reverse the slopes in figure~\ref{fig:charge_totals}. This is especially true if it is the case that the magnetic field, although smaller initially, has a longer lifetime at lower energies~\cite{Skokov:2009qp}. We leave these topics to be investigated in the future.

In fact, the many existing predictions for the evolution of the magnetic field in a heavy ion collision can be incorporated into our model and studied in the future. As one example, the magnetic field was found to be sustained longer than implied by $1/\tau$-behavior when the finite conductivity of the medium was taken into account~\cite{McLerran:2013hla}. 
Furthermore, because of combined effects of the electromagnetic response of the QGP and continued production of a field by the valence charges, the magnetic field was argued to be effectively constant during the lifetime of the QGP (from approximately $1\,{\rm fm}/c$ after the collision until freeze-out)~\cite{Stewart:2021mjz}. That calculation was based on a decoupling of the fluid dynamics and the dynamics of the electromagnetic fields~\cite{Stewart:2017zsu}. 
As another example, in a distinct approach the time-evolution and effects of electric and magnetic fields within proton-nucleus collisions at relativistic energy has been considered in the Parton-Hadron-
String Dynamics approach~\cite{Oliva:2019kin}, predicting a sizable electric field (while standard magnetohydrodynamic approaches neglect the electric field by definition). 
This situation of several distinct theoretical predictions for electromagnetic field behavior emphasizes the significant uncertainty stemming from our lack of detailed knowledge about the magnetic field; a situation which may be remedied considering these different scenarios within our model and comparing to experimental data such as the isobar run. 

We expect the effect of nonzero baryon or electric charge density to leave our results qualitatively unchanged based on previous non-linear~\cite{Cartwright2019} and linearized~\cite{Janiszewski:2015ura} computations in holographic models similar to ours.  

As a final note, we point out that there are two distinct pressure anisotropies which may serve as indicators for the time-evolution of the plasma and the magnetic field, respectively. 
If the system evolves according to viscous Bjorken flow, then the pressure anisotropy without a magnetic field is  $\Delta P = P_L-P_T\propto \tau^{-1}$. This is the anisotropy due to the viscosity created between the longitudinal pressure (along the rapidity direction defined with the beamline) and transverse plane pressure by the expansion along the beamline. If there is a magnetic field, then there is an additional anisotropy, now in the transverse plane. It is created between the pressure along the magnetic field $P_B$ and the pressure that is perpendicular to the magnetic field and also perpendicular to the beamline $P_{\perp\perp}$. If the magnetic field is constant over the lifetime of the plasma, then we expect the transverse plane anisotropy $\Delta P_\perp = P_B-P_{\perp\perp}$ to be constant over time. If the magnetic field decays as $B\propto \tau^{-1}$, then the $\Delta P_\perp\propto \tau^{-2}$. 
If these two anisotropies $\Delta P$ and $\Delta P_\perp$ could be measured, they would allow to distinguish between the magnetic field being constant, time-dependent, or absent during the lifetime of the plasma.\footnote{Our model does not include the potentially large anisotropy created by the initial geometry or fluctuations, although two of the authors have studied the time-evolution of that type of anisotropy within this framework before~\cite{Cartwright2019}.}


\acknowledgments
We thank Dmitri Kharzeev for comments as well as Marco Knipfer and Karl Landsteiner for discussions. C.C. and M.K. are supported, in part, by the U.S. Department of Energy grant DE-SC0012447 and B.P.S. is supported by the U.S. Department of Energy under Contract No.\,DE-SC0012704. The material presented in this article is based in part upon work done while C.C. was supported by the U.S. Department of Energy, Office of Science, Office of Workforce Development for Teachers and Scientists, Office of Science Graduate Student Research
(SCGSR) program. The SCGSR program is administered by the Oak Ridge Institute for Science and Education
(ORISE) for the DOE. ORISE is managed by ORAU under contract number DE‐SC0014664. All opinions expressed
in this paper are the author’s and do not necessarily reflect the policies and views of DOE, ORAU, or ORISE.

\bibliography{refCME.bib}

\begin{thebibliography}{70}%
\makeatletter
\providecommand \@ifxundefined [1]{%
 \@ifx{#1\undefined}
}%
\providecommand \@ifnum [1]{%
 \ifnum #1\expandafter \@firstoftwo
 \else \expandafter \@secondoftwo
 \fi
}%
\providecommand \@ifx [1]{%
 \ifx #1\expandafter \@firstoftwo
 \else \expandafter \@secondoftwo
 \fi
}%
\providecommand \natexlab [1]{#1}%
\providecommand \enquote  [1]{``#1''}%
\providecommand \bibnamefont  [1]{#1}%
\providecommand \bibfnamefont [1]{#1}%
\providecommand \citenamefont [1]{#1}%
\providecommand \href@noop [0]{\@secondoftwo}%
\providecommand \href [0]{\begingroup \@sanitize@url \@href}%
\providecommand \@href[1]{\@@startlink{#1}\@@href}%
\providecommand \@@href[1]{\endgroup#1\@@endlink}%
\providecommand \@sanitize@url [0]{\catcode `\\12\catcode `\$12\catcode
  `\&12\catcode `\#12\catcode `\^12\catcode `\_12\catcode `\%12\relax}%
\providecommand \@@startlink[1]{}%
\providecommand \@@endlink[0]{}%
\providecommand \url  [0]{\begingroup\@sanitize@url \@url }%
\providecommand \@url [1]{\endgroup\@href {#1}{\urlprefix }}%
\providecommand \urlprefix  [0]{URL }%
\providecommand \Eprint [0]{\href }%
\providecommand \doibase [0]{https://doi.org/}%
\providecommand \selectlanguage [0]{\@gobble}%
\providecommand \bibinfo  [0]{\@secondoftwo}%
\providecommand \bibfield  [0]{\@secondoftwo}%
\providecommand \translation [1]{[#1]}%
\providecommand \BibitemOpen [0]{}%
\providecommand \bibitemStop [0]{}%
\providecommand \bibitemNoStop [0]{.\EOS\space}%
\providecommand \EOS [0]{\spacefactor3000\relax}%
\providecommand \BibitemShut  [1]{\csname bibitem#1\endcsname}%
\let\auto@bib@innerbib\@empty
\bibitem [{\citenamefont {Kharzeev}(2006)}]{Kharzeev:2004ey}%
  \BibitemOpen
  \bibfield  {author} {\bibinfo {author} {\bibfnamefont {D.}~\bibnamefont
  {Kharzeev}},\ }\href {https://doi.org/10.1016/j.physletb.2005.11.075}
  {\bibfield  {journal} {\bibinfo  {journal} {Phys. Lett.}\ }\textbf {\bibinfo
  {volume} {B633}},\ \bibinfo {pages} {260} (\bibinfo {year} {2006})},\ \Eprint
  {https://arxiv.org/abs/hep-ph/0406125} {arXiv:hep-ph/0406125 [hep-ph]}
  \BibitemShut {NoStop}%
\bibitem [{\citenamefont {Kharzeev}\ \emph {et~al.}(2008)\citenamefont
  {Kharzeev}, \citenamefont {McLerran},\ and\ \citenamefont
  {Warringa}}]{Kharzeev:2007jp}%
  \BibitemOpen
  \bibfield  {author} {\bibinfo {author} {\bibfnamefont {D.~E.}\ \bibnamefont
  {Kharzeev}}, \bibinfo {author} {\bibfnamefont {L.~D.}\ \bibnamefont
  {McLerran}},\ and\ \bibinfo {author} {\bibfnamefont {H.~J.}\ \bibnamefont
  {Warringa}},\ }\href {https://doi.org/10.1016/j.nuclphysa.2008.02.298}
  {\bibfield  {journal} {\bibinfo  {journal} {Nucl. Phys.}\ }\textbf {\bibinfo
  {volume} {A803}},\ \bibinfo {pages} {227} (\bibinfo {year} {2008})},\ \Eprint
  {https://arxiv.org/abs/0711.0950} {arXiv:0711.0950 [hep-ph]} \BibitemShut
  {NoStop}%
\bibitem [{\citenamefont {Son}\ and\ \citenamefont
  {Surowka}(2009)}]{Son:2009tf}%
  \BibitemOpen
  \bibfield  {author} {\bibinfo {author} {\bibfnamefont {D.~T.}\ \bibnamefont
  {Son}}\ and\ \bibinfo {author} {\bibfnamefont {P.}~\bibnamefont {Surowka}},\
  }\href {https://doi.org/10.1103/PhysRevLett.103.191601} {\bibfield  {journal}
  {\bibinfo  {journal} {Phys. Rev. Lett.}\ }\textbf {\bibinfo {volume} {103}},\
  \bibinfo {pages} {191601} (\bibinfo {year} {2009})},\ \Eprint
  {https://arxiv.org/abs/0906.5044} {arXiv:0906.5044 [hep-th]} \BibitemShut
  {NoStop}%
\bibitem [{\citenamefont {Gynther}\ \emph {et~al.}(2011)\citenamefont
  {Gynther}, \citenamefont {Landsteiner}, \citenamefont {Pena-Benitez},\ and\
  \citenamefont {Rebhan}}]{Gynther:2010ed}%
  \BibitemOpen
  \bibfield  {author} {\bibinfo {author} {\bibfnamefont {A.}~\bibnamefont
  {Gynther}}, \bibinfo {author} {\bibfnamefont {K.}~\bibnamefont
  {Landsteiner}}, \bibinfo {author} {\bibfnamefont {F.}~\bibnamefont
  {Pena-Benitez}},\ and\ \bibinfo {author} {\bibfnamefont {A.}~\bibnamefont
  {Rebhan}},\ }\href {https://doi.org/10.1007/JHEP02(2011)110} {\bibfield
  {journal} {\bibinfo  {journal} {JHEP}\ }\textbf {\bibinfo {volume} {02}},\
  \bibinfo {pages} {110}},\ \Eprint {https://arxiv.org/abs/1005.2587}
  {arXiv:1005.2587 [hep-th]} \BibitemShut {NoStop}%
\bibitem [{\citenamefont {Amado}\ \emph {et~al.}(2011)\citenamefont {Amado},
  \citenamefont {Landsteiner},\ and\ \citenamefont
  {Pena-Benitez}}]{Amado:2011zx}%
  \BibitemOpen
  \bibfield  {author} {\bibinfo {author} {\bibfnamefont {I.}~\bibnamefont
  {Amado}}, \bibinfo {author} {\bibfnamefont {K.}~\bibnamefont {Landsteiner}},\
  and\ \bibinfo {author} {\bibfnamefont {F.}~\bibnamefont {Pena-Benitez}},\
  }\href {https://doi.org/10.1007/JHEP05(2011)081} {\bibfield  {journal}
  {\bibinfo  {journal} {JHEP}\ }\textbf {\bibinfo {volume} {05}},\ \bibinfo
  {pages} {081}},\ \Eprint {https://arxiv.org/abs/1102.4577} {arXiv:1102.4577
  [hep-th]} \BibitemShut {NoStop}%
\bibitem [{\citenamefont {Kharzeev}\ \emph {et~al.}(2013)\citenamefont
  {Kharzeev}, \citenamefont {Landsteiner}, \citenamefont {Schmitt},\ and\
  \citenamefont {Yee}}]{Kharzeev:2012ph}%
  \BibitemOpen
  \bibfield  {author} {\bibinfo {author} {\bibfnamefont {D.~E.}\ \bibnamefont
  {Kharzeev}}, \bibinfo {author} {\bibfnamefont {K.}~\bibnamefont
  {Landsteiner}}, \bibinfo {author} {\bibfnamefont {A.}~\bibnamefont
  {Schmitt}},\ and\ \bibinfo {author} {\bibfnamefont {H.-U.}\ \bibnamefont
  {Yee}},\ }\href {https://doi.org/10.1007/978-3-642-37305-3_1} {\bibfield
  {journal} {\bibinfo  {journal} {Lect. Notes Phys.}\ }\textbf {\bibinfo
  {volume} {871}},\ \bibinfo {pages} {1} (\bibinfo {year} {2013})},\ \Eprint
  {https://arxiv.org/abs/1211.6245} {arXiv:1211.6245 [hep-ph]} \BibitemShut
  {NoStop}%
\bibitem [{\citenamefont {Ammon}\ \emph {et~al.}(2021)\citenamefont {Ammon},
  \citenamefont {Grieninger}, \citenamefont {Hernandez}, \citenamefont
  {Kaminski}, \citenamefont {Koirala}, \citenamefont {Leiber},\ and\
  \citenamefont {Wu}}]{Ammon:2020rvg}%
  \BibitemOpen
  \bibfield  {author} {\bibinfo {author} {\bibfnamefont {M.}~\bibnamefont
  {Ammon}}, \bibinfo {author} {\bibfnamefont {S.}~\bibnamefont {Grieninger}},
  \bibinfo {author} {\bibfnamefont {J.}~\bibnamefont {Hernandez}}, \bibinfo
  {author} {\bibfnamefont {M.}~\bibnamefont {Kaminski}}, \bibinfo {author}
  {\bibfnamefont {R.}~\bibnamefont {Koirala}}, \bibinfo {author} {\bibfnamefont
  {J.}~\bibnamefont {Leiber}},\ and\ \bibinfo {author} {\bibfnamefont
  {J.}~\bibnamefont {Wu}},\ }\href {https://doi.org/10.1007/JHEP04(2021)078}
  {\bibfield  {journal} {\bibinfo  {journal} {JHEP}\ }\textbf {\bibinfo
  {volume} {04}},\ \bibinfo {pages} {078}},\ \Eprint
  {https://arxiv.org/abs/2012.09183} {arXiv:2012.09183 [hep-th]} \BibitemShut
  {NoStop}%
\bibitem [{\citenamefont {Ammon}\ \emph {et~al.}(2017)\citenamefont {Ammon},
  \citenamefont {Kaminski}, \citenamefont {Koirala}, \citenamefont {Leiber},\
  and\ \citenamefont {Wu}}]{Ammon:2017ded}%
  \BibitemOpen
  \bibfield  {author} {\bibinfo {author} {\bibfnamefont {M.}~\bibnamefont
  {Ammon}}, \bibinfo {author} {\bibfnamefont {M.}~\bibnamefont {Kaminski}},
  \bibinfo {author} {\bibfnamefont {R.}~\bibnamefont {Koirala}}, \bibinfo
  {author} {\bibfnamefont {J.}~\bibnamefont {Leiber}},\ and\ \bibinfo {author}
  {\bibfnamefont {J.}~\bibnamefont {Wu}},\ }\href@noop {} {\bibfield  {journal}
  {\bibinfo  {journal} {JHEP}\ }\textbf {\bibinfo {volume} {04}},\ \bibinfo
  {pages} {067}}\BibitemShut {NoStop}%
\bibitem [{\citenamefont {Bell}\ and\ \citenamefont
  {Jackiw}(1969)}]{Bell:1969ts}%
  \BibitemOpen
  \bibfield  {author} {\bibinfo {author} {\bibfnamefont {J.~S.}\ \bibnamefont
  {Bell}}\ and\ \bibinfo {author} {\bibfnamefont {R.}~\bibnamefont {Jackiw}},\
  }\href {https://doi.org/10.1007/BF02823296} {\bibfield  {journal} {\bibinfo
  {journal} {Nuovo Cim. A}\ }\textbf {\bibinfo {volume} {60}},\ \bibinfo
  {pages} {47} (\bibinfo {year} {1969})}\BibitemShut {NoStop}%
\bibitem [{\citenamefont {Adler}(1969)}]{Adler:1968av}%
  \BibitemOpen
  \bibfield  {author} {\bibinfo {author} {\bibfnamefont {S.~L.}\ \bibnamefont
  {Adler}},\ }\href {https://doi.org/10.1103/PhysRev.177.2426} {\bibfield
  {journal} {\bibinfo  {journal} {Phys. Rev.}\ }\textbf {\bibinfo {volume}
  {177}},\ \bibinfo {pages} {2426} (\bibinfo {year} {1969})}\BibitemShut
  {NoStop}%
\bibitem [{\citenamefont {Neiman}\ and\ \citenamefont
  {Oz}(2011)}]{Neiman:2010zi}%
  \BibitemOpen
  \bibfield  {author} {\bibinfo {author} {\bibfnamefont {Y.}~\bibnamefont
  {Neiman}}\ and\ \bibinfo {author} {\bibfnamefont {Y.}~\bibnamefont {Oz}},\
  }\href {https://doi.org/10.1007/JHEP03(2011)023} {\bibfield  {journal}
  {\bibinfo  {journal} {JHEP}\ }\textbf {\bibinfo {volume} {03}},\ \bibinfo
  {pages} {023}},\ \Eprint {https://arxiv.org/abs/1011.5107} {arXiv:1011.5107
  [hep-th]} \BibitemShut {NoStop}%
\bibitem [{\citenamefont {Son}\ and\ \citenamefont
  {Spivak}(2013)}]{Son:2012bg}%
  \BibitemOpen
  \bibfield  {author} {\bibinfo {author} {\bibfnamefont {D.~T.}\ \bibnamefont
  {Son}}\ and\ \bibinfo {author} {\bibfnamefont {B.~Z.}\ \bibnamefont
  {Spivak}},\ }\href {https://doi.org/10.1103/PhysRevB.88.104412} {\bibfield
  {journal} {\bibinfo  {journal} {Phys. Rev. B}\ }\textbf {\bibinfo {volume}
  {88}},\ \bibinfo {pages} {104412} (\bibinfo {year} {2013})},\ \Eprint
  {https://arxiv.org/abs/1206.1627} {arXiv:1206.1627 [cond-mat.mes-hall]}
  \BibitemShut {NoStop}%
\bibitem [{\citenamefont {Chernodub}\ \emph {et~al.}(2014)\citenamefont
  {Chernodub}, \citenamefont {Cortijo}, \citenamefont {Grushin}, \citenamefont
  {Landsteiner},\ and\ \citenamefont {Vozmediano}}]{Chernodub:2013kya}%
  \BibitemOpen
  \bibfield  {author} {\bibinfo {author} {\bibfnamefont {M.~N.}\ \bibnamefont
  {Chernodub}}, \bibinfo {author} {\bibfnamefont {A.}~\bibnamefont {Cortijo}},
  \bibinfo {author} {\bibfnamefont {A.~G.}\ \bibnamefont {Grushin}}, \bibinfo
  {author} {\bibfnamefont {K.}~\bibnamefont {Landsteiner}},\ and\ \bibinfo
  {author} {\bibfnamefont {M.~A.~H.}\ \bibnamefont {Vozmediano}},\ }\href
  {https://doi.org/10.1103/PhysRevB.89.081407} {\bibfield  {journal} {\bibinfo
  {journal} {Phys. Rev. B}\ }\textbf {\bibinfo {volume} {89}},\ \bibinfo
  {pages} {081407} (\bibinfo {year} {2014})},\ \Eprint
  {https://arxiv.org/abs/1311.0878} {arXiv:1311.0878 [hep-th]} \BibitemShut
  {NoStop}%
\bibitem [{\citenamefont {Cortijo}\ \emph {et~al.}(2015)\citenamefont
  {Cortijo}, \citenamefont {Ferreiros}, \citenamefont {Landsteiner},\ and\
  \citenamefont {Vozmediano}}]{Cortijo:2015hlt}%
  \BibitemOpen
  \bibfield  {author} {\bibinfo {author} {\bibfnamefont {A.}~\bibnamefont
  {Cortijo}}, \bibinfo {author} {\bibfnamefont {Y.}~\bibnamefont {Ferreiros}},
  \bibinfo {author} {\bibfnamefont {K.}~\bibnamefont {Landsteiner}},\ and\
  \bibinfo {author} {\bibfnamefont {M.~A.~H.}\ \bibnamefont {Vozmediano}},\
  }\href {https://doi.org/10.1103/PhysRevLett.115.177202} {\bibfield  {journal}
  {\bibinfo  {journal} {Phys. Rev. Lett.}\ }\textbf {\bibinfo {volume} {115}},\
  \bibinfo {pages} {177202} (\bibinfo {year} {2015})},\ \Eprint
  {https://arxiv.org/abs/1603.02674} {arXiv:1603.02674 [cond-mat.mes-hall]}
  \BibitemShut {NoStop}%
\bibitem [{\citenamefont {Landsteiner}\ and\ \citenamefont
  {Liu}(2016)}]{Landsteiner:2015lsa}%
  \BibitemOpen
  \bibfield  {author} {\bibinfo {author} {\bibfnamefont {K.}~\bibnamefont
  {Landsteiner}}\ and\ \bibinfo {author} {\bibfnamefont {Y.}~\bibnamefont
  {Liu}},\ }\href {https://doi.org/10.1016/j.physletb.2015.12.052} {\bibfield
  {journal} {\bibinfo  {journal} {Phys. Lett. B}\ }\textbf {\bibinfo {volume}
  {753}},\ \bibinfo {pages} {453} (\bibinfo {year} {2016})},\ \Eprint
  {https://arxiv.org/abs/1505.04772} {arXiv:1505.04772 [hep-th]} \BibitemShut
  {NoStop}%
\bibitem [{\citenamefont {Cortijo}\ \emph {et~al.}(2016)\citenamefont
  {Cortijo}, \citenamefont {Kharzeev}, \citenamefont {Landsteiner},\ and\
  \citenamefont {Vozmediano}}]{Cortijo:2016wnf}%
  \BibitemOpen
  \bibfield  {author} {\bibinfo {author} {\bibfnamefont {A.}~\bibnamefont
  {Cortijo}}, \bibinfo {author} {\bibfnamefont {D.}~\bibnamefont {Kharzeev}},
  \bibinfo {author} {\bibfnamefont {K.}~\bibnamefont {Landsteiner}},\ and\
  \bibinfo {author} {\bibfnamefont {M.~A.~H.}\ \bibnamefont {Vozmediano}},\
  }\href {https://doi.org/10.1103/PhysRevB.94.241405} {\bibfield  {journal}
  {\bibinfo  {journal} {Phys. Rev. B}\ }\textbf {\bibinfo {volume} {94}},\
  \bibinfo {pages} {241405} (\bibinfo {year} {2016})},\ \Eprint
  {https://arxiv.org/abs/1607.03491} {arXiv:1607.03491 [cond-mat.mes-hall]}
  \BibitemShut {NoStop}%
\bibitem [{\citenamefont {Li}\ \emph {et~al.}(2016{\natexlab{a}})\citenamefont
  {Li}, \citenamefont {Kharzeev}, \citenamefont {Zhang}, \citenamefont {Huang},
  \citenamefont {Pletikosic}, \citenamefont {Fedorov}, \citenamefont {Zhong},
  \citenamefont {Schneeloch}, \citenamefont {Gu},\ and\ \citenamefont
  {Valla}}]{Li:2014bha}%
  \BibitemOpen
  \bibfield  {author} {\bibinfo {author} {\bibfnamefont {Q.}~\bibnamefont
  {Li}}, \bibinfo {author} {\bibfnamefont {D.~E.}\ \bibnamefont {Kharzeev}},
  \bibinfo {author} {\bibfnamefont {C.}~\bibnamefont {Zhang}}, \bibinfo
  {author} {\bibfnamefont {Y.}~\bibnamefont {Huang}}, \bibinfo {author}
  {\bibfnamefont {I.}~\bibnamefont {Pletikosic}}, \bibinfo {author}
  {\bibfnamefont {A.~V.}\ \bibnamefont {Fedorov}}, \bibinfo {author}
  {\bibfnamefont {R.~D.}\ \bibnamefont {Zhong}}, \bibinfo {author}
  {\bibfnamefont {J.~A.}\ \bibnamefont {Schneeloch}}, \bibinfo {author}
  {\bibfnamefont {G.~D.}\ \bibnamefont {Gu}},\ and\ \bibinfo {author}
  {\bibfnamefont {T.}~\bibnamefont {Valla}},\ }\href
  {https://doi.org/10.1038/nphys3648} {\bibfield  {journal} {\bibinfo
  {journal} {Nature Phys.}\ }\textbf {\bibinfo {volume} {12}},\ \bibinfo
  {pages} {550} (\bibinfo {year} {2016}{\natexlab{a}})},\ \Eprint
  {https://arxiv.org/abs/1412.6543} {arXiv:1412.6543 [cond-mat.str-el]}
  \BibitemShut {NoStop}%
\bibitem [{\citenamefont {Arnold}\ \emph {et~al.}(2016)\citenamefont {Arnold}
  \emph {et~al.}}]{Arnold:2015vvs}%
  \BibitemOpen
  \bibfield  {author} {\bibinfo {author} {\bibfnamefont {F.}~\bibnamefont
  {Arnold}} \emph {et~al.},\ }\href {https://doi.org/10.1038/ncomms11615}
  {\bibfield  {journal} {\bibinfo  {journal} {Nature Commun.}\ }\textbf
  {\bibinfo {volume} {7}},\ \bibinfo {pages} {1615} (\bibinfo {year} {2016})},\
  \Eprint {https://arxiv.org/abs/1506.06577} {arXiv:1506.06577
  [cond-mat.mtrl-sci]} \BibitemShut {NoStop}%
\bibitem [{\citenamefont {Huang}\ \emph {et~al.}(2015)\citenamefont {Huang},
  \citenamefont {Zhao}, \citenamefont {Long}, \citenamefont {Wang},
  \citenamefont {Chen}, \citenamefont {Yang}, \citenamefont {Liang},
  \citenamefont {Xue}, \citenamefont {Weng}, \citenamefont {Fang},
  \citenamefont {Dai},\ and\ \citenamefont {Chen}}]{Huang:2015}%
  \BibitemOpen
  \bibfield  {author} {\bibinfo {author} {\bibfnamefont {X.}~\bibnamefont
  {Huang}}, \bibinfo {author} {\bibfnamefont {L.}~\bibnamefont {Zhao}},
  \bibinfo {author} {\bibfnamefont {Y.}~\bibnamefont {Long}}, \bibinfo {author}
  {\bibfnamefont {P.}~\bibnamefont {Wang}}, \bibinfo {author} {\bibfnamefont
  {D.}~\bibnamefont {Chen}}, \bibinfo {author} {\bibfnamefont {Z.}~\bibnamefont
  {Yang}}, \bibinfo {author} {\bibfnamefont {H.}~\bibnamefont {Liang}},
  \bibinfo {author} {\bibfnamefont {M.}~\bibnamefont {Xue}}, \bibinfo {author}
  {\bibfnamefont {H.}~\bibnamefont {Weng}}, \bibinfo {author} {\bibfnamefont
  {Z.}~\bibnamefont {Fang}}, \bibinfo {author} {\bibfnamefont {X.}~\bibnamefont
  {Dai}},\ and\ \bibinfo {author} {\bibfnamefont {G.}~\bibnamefont {Chen}},\
  }\href {https://doi.org/10.1103/PhysRevX.5.031023} {\bibfield  {journal}
  {\bibinfo  {journal} {Phys. Rev. X}\ }\textbf {\bibinfo {volume} {5}},\
  \bibinfo {pages} {031023} (\bibinfo {year} {2015})}\BibitemShut {NoStop}%
\bibitem [{\citenamefont {Xiong}\ \emph {et~al.}(2015)\citenamefont {Xiong},
  \citenamefont {Kushwaha}, \citenamefont {Liang}, \citenamefont {Krizan},
  \citenamefont {Hirschberger}, \citenamefont {Wang}, \citenamefont {Cava},\
  and\ \citenamefont {Ong}}]{Xiong:2015}%
  \BibitemOpen
  \bibfield  {author} {\bibinfo {author} {\bibfnamefont {J.}~\bibnamefont
  {Xiong}}, \bibinfo {author} {\bibfnamefont {S.~K.}\ \bibnamefont {Kushwaha}},
  \bibinfo {author} {\bibfnamefont {T.}~\bibnamefont {Liang}}, \bibinfo
  {author} {\bibfnamefont {J.~W.}\ \bibnamefont {Krizan}}, \bibinfo {author}
  {\bibfnamefont {M.}~\bibnamefont {Hirschberger}}, \bibinfo {author}
  {\bibfnamefont {W.}~\bibnamefont {Wang}}, \bibinfo {author} {\bibfnamefont
  {R.~J.}\ \bibnamefont {Cava}},\ and\ \bibinfo {author} {\bibfnamefont
  {N.~P.}\ \bibnamefont {Ong}},\ }\href
  {https://doi.org/10.1126/science.aac6089} {\bibfield  {journal} {\bibinfo
  {journal} {Science}\ }\textbf {\bibinfo {volume} {350}},\ \bibinfo {pages}
  {413} (\bibinfo {year} {2015})},\ \Eprint
  {https://arxiv.org/abs/https://www.science.org/doi/pdf/10.1126/science.aac6089}
  {https://www.science.org/doi/pdf/10.1126/science.aac6089} \BibitemShut
  {NoStop}%
\bibitem [{\citenamefont {Li}\ \emph {et~al.}(2015)\citenamefont {Li},
  \citenamefont {Wang}, \citenamefont {Liu}, \citenamefont {Wang},
  \citenamefont {Liao},\ and\ \citenamefont {Yu}}]{Li:2015}%
  \BibitemOpen
  \bibfield  {author} {\bibinfo {author} {\bibfnamefont {C.-Z.}\ \bibnamefont
  {Li}}, \bibinfo {author} {\bibfnamefont {L.-X.}\ \bibnamefont {Wang}},
  \bibinfo {author} {\bibfnamefont {H.}~\bibnamefont {Liu}}, \bibinfo {author}
  {\bibfnamefont {J.}~\bibnamefont {Wang}}, \bibinfo {author} {\bibfnamefont
  {Z.-M.}\ \bibnamefont {Liao}},\ and\ \bibinfo {author} {\bibfnamefont
  {D.-P.}\ \bibnamefont {Yu}},\ }\bibfield  {journal} {\bibinfo  {journal}
  {Nature Communications}\ }\textbf {\bibinfo {volume} {6}},\ \href
  {https://doi.org/10.1038/ncomms10137} {10.1038/ncomms10137} (\bibinfo {year}
  {2015})\BibitemShut {NoStop}%
\bibitem [{\citenamefont {Li}\ \emph {et~al.}(2016{\natexlab{b}})\citenamefont
  {Li}, \citenamefont {He}, \citenamefont {Lu}, \citenamefont {Zhang},
  \citenamefont {Liu}, \citenamefont {Ma}, \citenamefont {Fan}, \citenamefont
  {Shen},\ and\ \citenamefont {Wang}}]{Li:2016}%
  \BibitemOpen
  \bibfield  {author} {\bibinfo {author} {\bibfnamefont {H.}~\bibnamefont
  {Li}}, \bibinfo {author} {\bibfnamefont {H.}~\bibnamefont {He}}, \bibinfo
  {author} {\bibfnamefont {H.-Z.}\ \bibnamefont {Lu}}, \bibinfo {author}
  {\bibfnamefont {H.}~\bibnamefont {Zhang}}, \bibinfo {author} {\bibfnamefont
  {H.}~\bibnamefont {Liu}}, \bibinfo {author} {\bibfnamefont {R.}~\bibnamefont
  {Ma}}, \bibinfo {author} {\bibfnamefont {Z.}~\bibnamefont {Fan}}, \bibinfo
  {author} {\bibfnamefont {S.-Q.}\ \bibnamefont {Shen}},\ and\ \bibinfo
  {author} {\bibfnamefont {J.}~\bibnamefont {Wang}},\ }\bibfield  {journal}
  {\bibinfo  {journal} {Nature Communications}\ }\textbf {\bibinfo {volume}
  {7}},\ \href {https://doi.org/10.1038/ncomms10301} {10.1038/ncomms10301}
  (\bibinfo {year} {2016}{\natexlab{b}})\BibitemShut {NoStop}%
\bibitem [{\citenamefont {Hirschberger}\ \emph {et~al.}(2016)\citenamefont
  {Hirschberger}, \citenamefont {Kushwaha}, \citenamefont {Wang}, \citenamefont
  {Gibson}, \citenamefont {Liang}, \citenamefont {Belvin}, \citenamefont
  {Bernevig}, \citenamefont {Cava},\ and\ \citenamefont
  {Ong}}]{Hirschberger:2016}%
  \BibitemOpen
  \bibfield  {author} {\bibinfo {author} {\bibfnamefont {M.}~\bibnamefont
  {Hirschberger}}, \bibinfo {author} {\bibfnamefont {S.}~\bibnamefont
  {Kushwaha}}, \bibinfo {author} {\bibfnamefont {Z.}~\bibnamefont {Wang}},
  \bibinfo {author} {\bibfnamefont {Q.}~\bibnamefont {Gibson}}, \bibinfo
  {author} {\bibfnamefont {S.}~\bibnamefont {Liang}}, \bibinfo {author}
  {\bibfnamefont {C.~A.}\ \bibnamefont {Belvin}}, \bibinfo {author}
  {\bibfnamefont {B.~A.}\ \bibnamefont {Bernevig}}, \bibinfo {author}
  {\bibfnamefont {R.~J.}\ \bibnamefont {Cava}},\ and\ \bibinfo {author}
  {\bibfnamefont {N.~P.}\ \bibnamefont {Ong}},\ }\href
  {https://doi.org/10.1038/nmat4684} {\bibfield  {journal} {\bibinfo  {journal}
  {Nature Materials}\ }\textbf {\bibinfo {volume} {15}},\ \bibinfo {pages}
  {1161–1165} (\bibinfo {year} {2016})}\BibitemShut {NoStop}%
\bibitem [{\citenamefont {Gooth}\ \emph {et~al.}(2017)\citenamefont {Gooth}
  \emph {et~al.}}]{Gooth:2017mbd}%
  \BibitemOpen
  \bibfield  {author} {\bibinfo {author} {\bibfnamefont {J.}~\bibnamefont
  {Gooth}} \emph {et~al.},\ }\href {https://doi.org/10.1038/nature23005}
  {\bibfield  {journal} {\bibinfo  {journal} {Nature}\ }\textbf {\bibinfo
  {volume} {547}},\ \bibinfo {pages} {324} (\bibinfo {year} {2017})},\ \Eprint
  {https://arxiv.org/abs/1703.10682} {arXiv:1703.10682 [cond-mat.mtrl-sci]}
  \BibitemShut {NoStop}%
\bibitem [{\citenamefont {Shekhar}\ \emph {et~al.}(2018)\citenamefont
  {Shekhar}, \citenamefont {Kumar}, \citenamefont {Grinenko}, \citenamefont
  {Singh}, \citenamefont {Sarkar}, \citenamefont {Luetkens}, \citenamefont
  {Wu}, \citenamefont {Zhang}, \citenamefont {Komarek}, \citenamefont
  {Kampert},\ and\ \citenamefont {et~al.}}]{Shekhar:2018}%
  \BibitemOpen
  \bibfield  {author} {\bibinfo {author} {\bibfnamefont {C.}~\bibnamefont
  {Shekhar}}, \bibinfo {author} {\bibfnamefont {N.}~\bibnamefont {Kumar}},
  \bibinfo {author} {\bibfnamefont {V.}~\bibnamefont {Grinenko}}, \bibinfo
  {author} {\bibfnamefont {S.}~\bibnamefont {Singh}}, \bibinfo {author}
  {\bibfnamefont {R.}~\bibnamefont {Sarkar}}, \bibinfo {author} {\bibfnamefont
  {H.}~\bibnamefont {Luetkens}}, \bibinfo {author} {\bibfnamefont {S.-C.}\
  \bibnamefont {Wu}}, \bibinfo {author} {\bibfnamefont {Y.}~\bibnamefont
  {Zhang}}, \bibinfo {author} {\bibfnamefont {A.~C.}\ \bibnamefont {Komarek}},
  \bibinfo {author} {\bibfnamefont {E.}~\bibnamefont {Kampert}},\ and\ \bibinfo
  {author} {\bibnamefont {et~al.}},\ }\href
  {https://doi.org/10.1073/pnas.1810842115} {\bibfield  {journal} {\bibinfo
  {journal} {Proceedings of the National Academy of Sciences}\ }\textbf
  {\bibinfo {volume} {115}},\ \bibinfo {pages} {9140–9144} (\bibinfo {year}
  {2018})}\BibitemShut {NoStop}%
\bibitem [{\citenamefont {Abelev}\ \emph {et~al.}(2010)\citenamefont {Abelev}
  \emph {et~al.}}]{STAR:2009tro}%
  \BibitemOpen
  \bibfield  {author} {\bibinfo {author} {\bibfnamefont {B.~I.}\ \bibnamefont
  {Abelev}} \emph {et~al.} (\bibinfo {collaboration} {STAR}),\ }\href
  {https://doi.org/10.1103/PhysRevC.81.054908} {\bibfield  {journal} {\bibinfo
  {journal} {Phys. Rev. C}\ }\textbf {\bibinfo {volume} {81}},\ \bibinfo
  {pages} {054908} (\bibinfo {year} {2010})},\ \Eprint
  {https://arxiv.org/abs/0909.1717} {arXiv:0909.1717 [nucl-ex]} \BibitemShut
  {NoStop}%
\bibitem [{\citenamefont {Abelev}\ \emph {et~al.}(2009)\citenamefont {Abelev}
  \emph {et~al.}}]{STAR:2009wot}%
  \BibitemOpen
  \bibfield  {author} {\bibinfo {author} {\bibfnamefont {B.~I.}\ \bibnamefont
  {Abelev}} \emph {et~al.} (\bibinfo {collaboration} {STAR}),\ }\href
  {https://doi.org/10.1103/PhysRevLett.103.251601} {\bibfield  {journal}
  {\bibinfo  {journal} {Phys. Rev. Lett.}\ }\textbf {\bibinfo {volume} {103}},\
  \bibinfo {pages} {251601} (\bibinfo {year} {2009})},\ \Eprint
  {https://arxiv.org/abs/0909.1739} {arXiv:0909.1739 [nucl-ex]} \BibitemShut
  {NoStop}%
\bibitem [{\citenamefont {Adamczyk}\ \emph {et~al.}(2013)\citenamefont
  {Adamczyk} \emph {et~al.}}]{STAR:2013ksd}%
  \BibitemOpen
  \bibfield  {author} {\bibinfo {author} {\bibfnamefont {L.}~\bibnamefont
  {Adamczyk}} \emph {et~al.} (\bibinfo {collaboration} {STAR}),\ }\href
  {https://doi.org/10.1103/PhysRevC.88.064911} {\bibfield  {journal} {\bibinfo
  {journal} {Phys. Rev. C}\ }\textbf {\bibinfo {volume} {88}},\ \bibinfo
  {pages} {064911} (\bibinfo {year} {2013})},\ \Eprint
  {https://arxiv.org/abs/1302.3802} {arXiv:1302.3802 [nucl-ex]} \BibitemShut
  {NoStop}%
\bibitem [{\citenamefont {Adamczyk}\ \emph
  {et~al.}(2014{\natexlab{a}})\citenamefont {Adamczyk} \emph
  {et~al.}}]{STAR:2013zgu}%
  \BibitemOpen
  \bibfield  {author} {\bibinfo {author} {\bibfnamefont {L.}~\bibnamefont
  {Adamczyk}} \emph {et~al.} (\bibinfo {collaboration} {STAR}),\ }\href
  {https://doi.org/10.1103/PhysRevC.89.044908} {\bibfield  {journal} {\bibinfo
  {journal} {Phys. Rev. C}\ }\textbf {\bibinfo {volume} {89}},\ \bibinfo
  {pages} {044908} (\bibinfo {year} {2014}{\natexlab{a}})},\ \Eprint
  {https://arxiv.org/abs/1303.0901} {arXiv:1303.0901 [nucl-ex]} \BibitemShut
  {NoStop}%
\bibitem [{\citenamefont {Adamczyk}\ \emph
  {et~al.}(2014{\natexlab{b}})\citenamefont {Adamczyk} \emph
  {et~al.}}]{STAR:2014uiw}%
  \BibitemOpen
  \bibfield  {author} {\bibinfo {author} {\bibfnamefont {L.}~\bibnamefont
  {Adamczyk}} \emph {et~al.} (\bibinfo {collaboration} {STAR}),\ }\href
  {https://doi.org/10.1103/PhysRevLett.113.052302} {\bibfield  {journal}
  {\bibinfo  {journal} {Phys. Rev. Lett.}\ }\textbf {\bibinfo {volume} {113}},\
  \bibinfo {pages} {052302} (\bibinfo {year} {2014}{\natexlab{b}})},\ \Eprint
  {https://arxiv.org/abs/1404.1433} {arXiv:1404.1433 [nucl-ex]} \BibitemShut
  {NoStop}%
\bibitem [{\citenamefont {Adam}\ \emph {et~al.}(2019)\citenamefont {Adam} \emph
  {et~al.}}]{STAR:2019xzd}%
  \BibitemOpen
  \bibfield  {author} {\bibinfo {author} {\bibfnamefont {J.}~\bibnamefont
  {Adam}} \emph {et~al.} (\bibinfo {collaboration} {STAR}),\ }\href
  {https://doi.org/10.1016/j.physletb.2019.134975} {\bibfield  {journal}
  {\bibinfo  {journal} {Phys. Lett. B}\ }\textbf {\bibinfo {volume} {798}},\
  \bibinfo {pages} {134975} (\bibinfo {year} {2019})},\ \Eprint
  {https://arxiv.org/abs/1906.03373} {arXiv:1906.03373 [nucl-ex]} \BibitemShut
  {NoStop}%
\bibitem [{\citenamefont {Abdallah}\ \emph {et~al.}(2021)\citenamefont
  {Abdallah} \emph {et~al.}}]{STAR:2021mii}%
  \BibitemOpen
  \bibfield  {author} {\bibinfo {author} {\bibfnamefont {M.}~\bibnamefont
  {Abdallah}} \emph {et~al.} (\bibinfo {collaboration} {STAR}),\ }\href@noop {}
  {\  (\bibinfo {year} {2021})},\ \Eprint {https://arxiv.org/abs/2109.00131}
  {arXiv:2109.00131 [nucl-ex]} \BibitemShut {NoStop}%
\bibitem [{\citenamefont {Abelev}\ \emph {et~al.}(2013)\citenamefont {Abelev}
  \emph {et~al.}}]{ALICE:2012nhw}%
  \BibitemOpen
  \bibfield  {author} {\bibinfo {author} {\bibfnamefont {B.}~\bibnamefont
  {Abelev}} \emph {et~al.} (\bibinfo {collaboration} {ALICE}),\ }\href
  {https://doi.org/10.1103/PhysRevLett.110.012301} {\bibfield  {journal}
  {\bibinfo  {journal} {Phys. Rev. Lett.}\ }\textbf {\bibinfo {volume} {110}},\
  \bibinfo {pages} {012301} (\bibinfo {year} {2013})},\ \Eprint
  {https://arxiv.org/abs/1207.0900} {arXiv:1207.0900 [nucl-ex]} \BibitemShut
  {NoStop}%
\bibitem [{\citenamefont {Acharya}\ \emph {et~al.}(2018)\citenamefont {Acharya}
  \emph {et~al.}}]{ALICE:2017sss}%
  \BibitemOpen
  \bibfield  {author} {\bibinfo {author} {\bibfnamefont {S.}~\bibnamefont
  {Acharya}} \emph {et~al.} (\bibinfo {collaboration} {ALICE}),\ }\href
  {https://doi.org/10.1016/j.physletb.2017.12.021} {\bibfield  {journal}
  {\bibinfo  {journal} {Phys. Lett. B}\ }\textbf {\bibinfo {volume} {777}},\
  \bibinfo {pages} {151} (\bibinfo {year} {2018})},\ \Eprint
  {https://arxiv.org/abs/1709.04723} {arXiv:1709.04723 [nucl-ex]} \BibitemShut
  {NoStop}%
\bibitem [{\citenamefont {Acharya}\ \emph {et~al.}(2020)\citenamefont {Acharya}
  \emph {et~al.}}]{ALICE:2020siw}%
  \BibitemOpen
  \bibfield  {author} {\bibinfo {author} {\bibfnamefont {S.}~\bibnamefont
  {Acharya}} \emph {et~al.} (\bibinfo {collaboration} {ALICE}),\ }\href
  {https://doi.org/10.1007/JHEP09(2020)160} {\bibfield  {journal} {\bibinfo
  {journal} {JHEP}\ }\textbf {\bibinfo {volume} {09}},\ \bibinfo {pages}
  {160}},\ \Eprint {https://arxiv.org/abs/2005.14640} {arXiv:2005.14640
  [nucl-ex]} \BibitemShut {NoStop}%
\bibitem [{\citenamefont {Khachatryan}\ \emph {et~al.}(2017)\citenamefont
  {Khachatryan} \emph {et~al.}}]{CMS:2016wfo}%
  \BibitemOpen
  \bibfield  {author} {\bibinfo {author} {\bibfnamefont {V.}~\bibnamefont
  {Khachatryan}} \emph {et~al.} (\bibinfo {collaboration} {CMS}),\ }\href
  {https://doi.org/10.1103/PhysRevLett.118.122301} {\bibfield  {journal}
  {\bibinfo  {journal} {Phys. Rev. Lett.}\ }\textbf {\bibinfo {volume} {118}},\
  \bibinfo {pages} {122301} (\bibinfo {year} {2017})},\ \Eprint
  {https://arxiv.org/abs/1610.00263} {arXiv:1610.00263 [nucl-ex]} \BibitemShut
  {NoStop}%
\bibitem [{\citenamefont {Sirunyan}\ \emph {et~al.}(2018)\citenamefont
  {Sirunyan} \emph {et~al.}}]{CMS:2017lrw}%
  \BibitemOpen
  \bibfield  {author} {\bibinfo {author} {\bibfnamefont {A.~M.}\ \bibnamefont
  {Sirunyan}} \emph {et~al.} (\bibinfo {collaboration} {CMS}),\ }\href
  {https://doi.org/10.1103/PhysRevC.97.044912} {\bibfield  {journal} {\bibinfo
  {journal} {Phys. Rev. C}\ }\textbf {\bibinfo {volume} {97}},\ \bibinfo
  {pages} {044912} (\bibinfo {year} {2018})},\ \Eprint
  {https://arxiv.org/abs/1708.01602} {arXiv:1708.01602 [nucl-ex]} \BibitemShut
  {NoStop}%
\bibitem [{\citenamefont {Marr}\ \emph {et~al.}(2019)\citenamefont {Marr} \emph
  {et~al.}}]{Marr:IPAC2019-MOZPLS2}%
  \BibitemOpen
  \bibfield  {author} {\bibinfo {author} {\bibfnamefont {G.}~\bibnamefont
  {Marr}} \emph {et~al.},\ }in\ \href
  {https://doi.org/doi:10.18429/JACoW-IPAC2019-MOZPLS2} {\emph {\bibinfo
  {booktitle} {Proc. 10th International Particle Accelerator Conference
  (IPAC'19), Melbourne, Australia, 19-24 May 2019}}},\ \bibinfo {series and
  number} {\bibinfo {series} {International Particle Accelerator Conference}\
  No.~\bibinfo {number} {10}}\ (\bibinfo  {publisher} {JACoW Publishing},\
  \bibinfo {address} {Geneva, Switzerland},\ \bibinfo {year} {2019})\ pp.\
  \bibinfo {pages} {28--32},\ \bibinfo {note}
  {https://doi.org/10.18429/JACoW-IPAC2019-MOZPLS2}\BibitemShut {NoStop}%
\bibitem [{\citenamefont {Deng}\ and\ \citenamefont
  {Huang}(2012)}]{Deng:2012pc}%
  \BibitemOpen
  \bibfield  {author} {\bibinfo {author} {\bibfnamefont {W.-T.}\ \bibnamefont
  {Deng}}\ and\ \bibinfo {author} {\bibfnamefont {X.-G.}\ \bibnamefont
  {Huang}},\ }\href {https://doi.org/10.1103/PhysRevC.85.044907} {\bibfield
  {journal} {\bibinfo  {journal} {Phys. Rev. C}\ }\textbf {\bibinfo {volume}
  {85}},\ \bibinfo {pages} {044907} (\bibinfo {year} {2012})},\ \Eprint
  {https://arxiv.org/abs/1201.5108} {arXiv:1201.5108 [nucl-th]} \BibitemShut
  {NoStop}%
\bibitem [{\citenamefont {McLerran}\ and\ \citenamefont
  {Skokov}(2014)}]{McLerran:2013hla}%
  \BibitemOpen
  \bibfield  {author} {\bibinfo {author} {\bibfnamefont {L.}~\bibnamefont
  {McLerran}}\ and\ \bibinfo {author} {\bibfnamefont {V.}~\bibnamefont
  {Skokov}},\ }\href {https://doi.org/10.1016/j.nuclphysa.2014.05.008}
  {\bibfield  {journal} {\bibinfo  {journal} {Nucl. Phys. A}\ }\textbf
  {\bibinfo {volume} {929}},\ \bibinfo {pages} {184} (\bibinfo {year}
  {2014})},\ \Eprint {https://arxiv.org/abs/1305.0774} {arXiv:1305.0774
  [hep-ph]} \BibitemShut {NoStop}%
\bibitem [{\citenamefont {Roy}\ \emph {et~al.}(2015)\citenamefont {Roy},
  \citenamefont {Pu}, \citenamefont {Rezzolla},\ and\ \citenamefont
  {Rischke}}]{Roy:2015kma}%
  \BibitemOpen
  \bibfield  {author} {\bibinfo {author} {\bibfnamefont {V.}~\bibnamefont
  {Roy}}, \bibinfo {author} {\bibfnamefont {S.}~\bibnamefont {Pu}}, \bibinfo
  {author} {\bibfnamefont {L.}~\bibnamefont {Rezzolla}},\ and\ \bibinfo
  {author} {\bibfnamefont {D.}~\bibnamefont {Rischke}},\ }\href
  {https://doi.org/10.1016/j.physletb.2015.08.046} {\bibfield  {journal}
  {\bibinfo  {journal} {Phys. Lett. B}\ }\textbf {\bibinfo {volume} {750}},\
  \bibinfo {pages} {45} (\bibinfo {year} {2015})},\ \Eprint
  {https://arxiv.org/abs/1506.06620} {arXiv:1506.06620 [nucl-th]} \BibitemShut
  {NoStop}%
\bibitem [{\citenamefont {Pu}\ \emph {et~al.}(2016)\citenamefont {Pu},
  \citenamefont {Roy}, \citenamefont {Rezzolla},\ and\ \citenamefont
  {Rischke}}]{Pu:2016ayh}%
  \BibitemOpen
  \bibfield  {author} {\bibinfo {author} {\bibfnamefont {S.}~\bibnamefont
  {Pu}}, \bibinfo {author} {\bibfnamefont {V.}~\bibnamefont {Roy}}, \bibinfo
  {author} {\bibfnamefont {L.}~\bibnamefont {Rezzolla}},\ and\ \bibinfo
  {author} {\bibfnamefont {D.~H.}\ \bibnamefont {Rischke}},\ }\href
  {https://doi.org/10.1103/PhysRevD.93.074022} {\bibfield  {journal} {\bibinfo
  {journal} {Phys. Rev. D}\ }\textbf {\bibinfo {volume} {93}},\ \bibinfo
  {pages} {074022} (\bibinfo {year} {2016})},\ \Eprint
  {https://arxiv.org/abs/1602.04953} {arXiv:1602.04953 [nucl-th]} \BibitemShut
  {NoStop}%
\bibitem [{\citenamefont {Stewart}\ and\ \citenamefont
  {Tuchin}(2018)}]{Stewart:2017zsu}%
  \BibitemOpen
  \bibfield  {author} {\bibinfo {author} {\bibfnamefont {E.}~\bibnamefont
  {Stewart}}\ and\ \bibinfo {author} {\bibfnamefont {K.}~\bibnamefont
  {Tuchin}},\ }\href {https://doi.org/10.1103/PhysRevC.97.044906} {\bibfield
  {journal} {\bibinfo  {journal} {Phys. Rev. C}\ }\textbf {\bibinfo {volume}
  {97}},\ \bibinfo {pages} {044906} (\bibinfo {year} {2018})},\ \Eprint
  {https://arxiv.org/abs/1710.08793} {arXiv:1710.08793 [nucl-th]} \BibitemShut
  {NoStop}%
\bibitem [{\citenamefont {Oliva}\ \emph {et~al.}(2020)\citenamefont {Oliva},
  \citenamefont {Moreau}, \citenamefont {Voronyuk},\ and\ \citenamefont
  {Bratkovskaya}}]{Oliva:2019kin}%
  \BibitemOpen
  \bibfield  {author} {\bibinfo {author} {\bibfnamefont {L.}~\bibnamefont
  {Oliva}}, \bibinfo {author} {\bibfnamefont {P.}~\bibnamefont {Moreau}},
  \bibinfo {author} {\bibfnamefont {V.}~\bibnamefont {Voronyuk}},\ and\
  \bibinfo {author} {\bibfnamefont {E.}~\bibnamefont {Bratkovskaya}},\ }\href
  {https://doi.org/10.1103/PhysRevC.101.014917} {\bibfield  {journal} {\bibinfo
   {journal} {Phys. Rev. C}\ }\textbf {\bibinfo {volume} {101}},\ \bibinfo
  {pages} {014917} (\bibinfo {year} {2020})},\ \Eprint
  {https://arxiv.org/abs/1909.06770} {arXiv:1909.06770 [nucl-th]} \BibitemShut
  {NoStop}%
\bibitem [{\citenamefont {Yan}\ and\ \citenamefont
  {Huang}(2021)}]{Yan:2021zjc}%
  \BibitemOpen
  \bibfield  {author} {\bibinfo {author} {\bibfnamefont {L.}~\bibnamefont
  {Yan}}\ and\ \bibinfo {author} {\bibfnamefont {X.-G.}\ \bibnamefont
  {Huang}},\ }\href@noop {} {\  (\bibinfo {year} {2021})},\ \Eprint
  {https://arxiv.org/abs/2104.00831} {arXiv:2104.00831 [nucl-th]} \BibitemShut
  {NoStop}%
\bibitem [{\citenamefont {Stewart}\ and\ \citenamefont
  {Tuchin}(2021)}]{Stewart:2021mjz}%
  \BibitemOpen
  \bibfield  {author} {\bibinfo {author} {\bibfnamefont {E.}~\bibnamefont
  {Stewart}}\ and\ \bibinfo {author} {\bibfnamefont {K.}~\bibnamefont
  {Tuchin}},\ }\href@noop {} {\  (\bibinfo {year} {2021})},\ \Eprint
  {https://arxiv.org/abs/2106.09124} {arXiv:2106.09124 [nucl-th]} \BibitemShut
  {NoStop}%
\bibitem [{\citenamefont {Siddique}\ \emph {et~al.}(2019)\citenamefont
  {Siddique}, \citenamefont {Wang}, \citenamefont {Pu},\ and\ \citenamefont
  {Wang}}]{Siddique:2019gqh}%
  \BibitemOpen
  \bibfield  {author} {\bibinfo {author} {\bibfnamefont {I.}~\bibnamefont
  {Siddique}}, \bibinfo {author} {\bibfnamefont {R.-j.}\ \bibnamefont {Wang}},
  \bibinfo {author} {\bibfnamefont {S.}~\bibnamefont {Pu}},\ and\ \bibinfo
  {author} {\bibfnamefont {Q.}~\bibnamefont {Wang}},\ }\href
  {https://doi.org/10.1103/PhysRevD.99.114029} {\bibfield  {journal} {\bibinfo
  {journal} {Phys. Rev. D}\ }\textbf {\bibinfo {volume} {99}},\ \bibinfo
  {pages} {114029} (\bibinfo {year} {2019})},\ \Eprint
  {https://arxiv.org/abs/1904.01807} {arXiv:1904.01807 [hep-ph]} \BibitemShut
  {NoStop}%
\bibitem [{\citenamefont {Ghosh}\ \emph {et~al.}(2021)\citenamefont {Ghosh},
  \citenamefont {Grieninger}, \citenamefont {Landsteiner},\ and\ \citenamefont
  {Morales-Tejera}}]{Ghosh:2021naw}%
  \BibitemOpen
  \bibfield  {author} {\bibinfo {author} {\bibfnamefont {J.~K.}\ \bibnamefont
  {Ghosh}}, \bibinfo {author} {\bibfnamefont {S.}~\bibnamefont {Grieninger}},
  \bibinfo {author} {\bibfnamefont {K.}~\bibnamefont {Landsteiner}},\ and\
  \bibinfo {author} {\bibfnamefont {S.}~\bibnamefont {Morales-Tejera}},\ }\href
  {https://doi.org/10.1103/PhysRevD.104.046009} {\bibfield  {journal} {\bibinfo
   {journal} {Phys. Rev. D}\ }\textbf {\bibinfo {volume} {104}},\ \bibinfo
  {pages} {046009} (\bibinfo {year} {2021})},\ \Eprint
  {https://arxiv.org/abs/2105.05855} {arXiv:2105.05855 [hep-ph]} \BibitemShut
  {NoStop}%
\bibitem [{\citenamefont {Taylor}(2000)}]{Taylor:2000xw}%
  \BibitemOpen
  \bibfield  {author} {\bibinfo {author} {\bibfnamefont {M.}~\bibnamefont
  {Taylor}},\ }\href@noop {} {\  (\bibinfo {year} {2000})},\ \Eprint
  {https://arxiv.org/abs/hep-th/0002125} {arXiv:hep-th/0002125 [hep-th]}
  \BibitemShut {NoStop}%
\bibitem [{\citenamefont {Chesler}\ and\ \citenamefont
  {Yaffe}(2009)}]{Chesler:2008hg}%
  \BibitemOpen
  \bibfield  {author} {\bibinfo {author} {\bibfnamefont {P.~M.}\ \bibnamefont
  {Chesler}}\ and\ \bibinfo {author} {\bibfnamefont {L.~G.}\ \bibnamefont
  {Yaffe}},\ }\href {https://doi.org/10.1103/PhysRevLett.102.211601} {\bibfield
   {journal} {\bibinfo  {journal} {Phys. Rev. Lett.}\ }\textbf {\bibinfo
  {volume} {102}},\ \bibinfo {pages} {211601} (\bibinfo {year} {2009})},\
  \Eprint {https://arxiv.org/abs/0812.2053} {arXiv:0812.2053 [hep-th]}
  \BibitemShut {NoStop}%
\bibitem [{\citenamefont {Cartwright}(2021)}]{Cartwright:2020qov}%
  \BibitemOpen
  \bibfield  {author} {\bibinfo {author} {\bibfnamefont {C.}~\bibnamefont
  {Cartwright}},\ }\href {https://doi.org/10.1007/JHEP01(2021)041} {\bibfield
  {journal} {\bibinfo  {journal} {JHEP}\ }\textbf {\bibinfo {volume} {01}},\
  \bibinfo {pages} {041}},\ \Eprint {https://arxiv.org/abs/2003.04325}
  {arXiv:2003.04325 [hep-th]} \BibitemShut {NoStop}%
\bibitem [{\citenamefont {Chesler}\ and\ \citenamefont
  {Yaffe}(2014)}]{Chesler:2013lia}%
  \BibitemOpen
  \bibfield  {author} {\bibinfo {author} {\bibfnamefont {P.~M.}\ \bibnamefont
  {Chesler}}\ and\ \bibinfo {author} {\bibfnamefont {L.~G.}\ \bibnamefont
  {Yaffe}},\ }\href {https://doi.org/10.1007/JHEP07(2014)086} {\bibfield
  {journal} {\bibinfo  {journal} {JHEP}\ }\textbf {\bibinfo {volume} {07}},\
  \bibinfo {pages} {086}},\ \Eprint {https://arxiv.org/abs/1309.1439}
  {arXiv:1309.1439 [hep-th]} \BibitemShut {NoStop}%
\bibitem [{\citenamefont {Skenderis}\ and\ \citenamefont {van
  Rees}(2009)}]{Skenderis:2008dg}%
  \BibitemOpen
  \bibfield  {author} {\bibinfo {author} {\bibfnamefont {K.}~\bibnamefont
  {Skenderis}}\ and\ \bibinfo {author} {\bibfnamefont {B.~C.}\ \bibnamefont
  {van Rees}},\ }\href {https://doi.org/10.1088/1126-6708/2009/05/085}
  {\bibfield  {journal} {\bibinfo  {journal} {JHEP}\ }\textbf {\bibinfo
  {volume} {05}},\ \bibinfo {pages} {085}},\ \Eprint
  {https://arxiv.org/abs/0812.2909} {arXiv:0812.2909 [hep-th]} \BibitemShut
  {NoStop}%
\bibitem [{\citenamefont {Fuini}\ and\ \citenamefont
  {Yaffe}(2015)}]{Fuini:2015hba}%
  \BibitemOpen
  \bibfield  {author} {\bibinfo {author} {\bibfnamefont {J.~F.}\ \bibnamefont
  {Fuini}}\ and\ \bibinfo {author} {\bibfnamefont {L.~G.}\ \bibnamefont
  {Yaffe}},\ }\href {https://doi.org/10.1007/JHEP07(2015)116} {\bibfield
  {journal} {\bibinfo  {journal} {JHEP}\ }\textbf {\bibinfo {volume} {07}},\
  \bibinfo {pages} {116}},\ \Eprint {https://arxiv.org/abs/1503.07148}
  {arXiv:1503.07148 [hep-th]} \BibitemShut {NoStop}%
\bibitem [{\citenamefont {Fern\'andez-Pend\'as}\ and\ \citenamefont
  {Landsteiner}(2019)}]{Pendas:2019}%
  \BibitemOpen
  \bibfield  {author} {\bibinfo {author} {\bibfnamefont {J.}~\bibnamefont
  {Fern\'andez-Pend\'as}}\ and\ \bibinfo {author} {\bibfnamefont
  {K.}~\bibnamefont {Landsteiner}},\ }\href
  {https://doi.org/10.1103/PhysRevD.100.126024} {\bibfield  {journal} {\bibinfo
   {journal} {Phys. Rev. D}\ }\textbf {\bibinfo {volume} {100}},\ \bibinfo
  {pages} {126024} (\bibinfo {year} {2019})}\BibitemShut {NoStop}%
\bibitem [{\citenamefont {Bardeen}\ and\ \citenamefont
  {Zumino}(1984)}]{Bardeen:1984pm}%
  \BibitemOpen
  \bibfield  {author} {\bibinfo {author} {\bibfnamefont {W.~A.}\ \bibnamefont
  {Bardeen}}\ and\ \bibinfo {author} {\bibfnamefont {B.}~\bibnamefont
  {Zumino}},\ }\href {https://doi.org/10.1016/0550-3213(84)90322-5} {\bibfield
  {journal} {\bibinfo  {journal} {Nucl. Phys. B}\ }\textbf {\bibinfo {volume}
  {244}},\ \bibinfo {pages} {421} (\bibinfo {year} {1984})}\BibitemShut
  {NoStop}%
\bibitem [{\citenamefont {Landsteiner}(2016)}]{Landsteiner:2016led}%
  \BibitemOpen
  \bibfield  {author} {\bibinfo {author} {\bibfnamefont {K.}~\bibnamefont
  {Landsteiner}},\ }\href {https://doi.org/10.5506/APhysPolB.47.2617}
  {\bibfield  {journal} {\bibinfo  {journal} {Acta Phys. Polon. B}\ }\textbf
  {\bibinfo {volume} {47}},\ \bibinfo {pages} {2617} (\bibinfo {year}
  {2016})},\ \Eprint {https://arxiv.org/abs/1610.04413} {arXiv:1610.04413
  [hep-th]} \BibitemShut {NoStop}%
\bibitem [{\citenamefont {Critelli}\ \emph {et~al.}(2018)\citenamefont
  {Critelli}, \citenamefont {Rougemont},\ and\ \citenamefont
  {Noronha}}]{Critelli:2018osu}%
  \BibitemOpen
  \bibfield  {author} {\bibinfo {author} {\bibfnamefont {R.}~\bibnamefont
  {Critelli}}, \bibinfo {author} {\bibfnamefont {R.}~\bibnamefont
  {Rougemont}},\ and\ \bibinfo {author} {\bibfnamefont {J.}~\bibnamefont
  {Noronha}},\ }\href@noop {} {\  (\bibinfo {year} {2018})},\ \Eprint
  {https://arxiv.org/abs/1805.00882} {arXiv:1805.00882 [hep-th]} \BibitemShut
  {NoStop}%
\bibitem [{\citenamefont {Rougemont}\ \emph {et~al.}(2021)\citenamefont
  {Rougemont}, \citenamefont {Noronha}, \citenamefont {Barreto}, \citenamefont
  {Denicol},\ and\ \citenamefont {Dore}}]{Rougemont:2021qyk}%
  \BibitemOpen
  \bibfield  {author} {\bibinfo {author} {\bibfnamefont {R.}~\bibnamefont
  {Rougemont}}, \bibinfo {author} {\bibfnamefont {J.}~\bibnamefont {Noronha}},
  \bibinfo {author} {\bibfnamefont {W.}~\bibnamefont {Barreto}}, \bibinfo
  {author} {\bibfnamefont {G.~S.}\ \bibnamefont {Denicol}},\ and\ \bibinfo
  {author} {\bibfnamefont {T.}~\bibnamefont {Dore}},\ }\href@noop {} {\
  (\bibinfo {year} {2021})},\ \Eprint {https://arxiv.org/abs/2105.02378}
  {arXiv:2105.02378 [nucl-th]} \BibitemShut {NoStop}%
\bibitem [{\citenamefont {Gubser}\ \emph {et~al.}(1996)\citenamefont {Gubser},
  \citenamefont {Klebanov},\ and\ \citenamefont {Peet}}]{Gubser:1996de}%
  \BibitemOpen
  \bibfield  {author} {\bibinfo {author} {\bibfnamefont {S.}~\bibnamefont
  {Gubser}}, \bibinfo {author} {\bibfnamefont {I.~R.}\ \bibnamefont
  {Klebanov}},\ and\ \bibinfo {author} {\bibfnamefont {A.}~\bibnamefont
  {Peet}},\ }\href {https://doi.org/10.1103/PhysRevD.54.3915} {\bibfield
  {journal} {\bibinfo  {journal} {Phys. Rev. D}\ }\textbf {\bibinfo {volume}
  {54}},\ \bibinfo {pages} {3915} (\bibinfo {year} {1996})},\ \Eprint
  {https://arxiv.org/abs/hep-th/9602135} {arXiv:hep-th/9602135} \BibitemShut
  {NoStop}%
\bibitem [{\citenamefont {Teper}(1999)}]{Teper:1998te}%
  \BibitemOpen
  \bibfield  {author} {\bibinfo {author} {\bibfnamefont {M.~J.}\ \bibnamefont
  {Teper}},\ }\href {https://doi.org/10.1103/PhysRevD.59.014512} {\bibfield
  {journal} {\bibinfo  {journal} {Phys. Rev. D}\ }\textbf {\bibinfo {volume}
  {59}},\ \bibinfo {pages} {014512} (\bibinfo {year} {1999})},\ \Eprint
  {https://arxiv.org/abs/hep-lat/9804008} {arXiv:hep-lat/9804008} \BibitemShut
  {NoStop}%
\bibitem [{\citenamefont {Meyer}\ and\ \citenamefont
  {Teper}(2003)}]{Meyer:2003wx}%
  \BibitemOpen
  \bibfield  {author} {\bibinfo {author} {\bibfnamefont {H.~B.}\ \bibnamefont
  {Meyer}}\ and\ \bibinfo {author} {\bibfnamefont {M.~J.}\ \bibnamefont
  {Teper}},\ }\href {https://doi.org/10.1016/j.nuclphysb.2003.07.003}
  {\bibfield  {journal} {\bibinfo  {journal} {Nucl. Phys. B}\ }\textbf
  {\bibinfo {volume} {668}},\ \bibinfo {pages} {111} (\bibinfo {year}
  {2003})},\ \Eprint {https://arxiv.org/abs/hep-lat/0306019}
  {arXiv:hep-lat/0306019} \BibitemShut {NoStop}%
\bibitem [{\citenamefont {Panero}(2012)}]{Panero:2012qx}%
  \BibitemOpen
  \bibfield  {author} {\bibinfo {author} {\bibfnamefont {M.}~\bibnamefont
  {Panero}},\ }\href {https://doi.org/10.22323/1.164.0010} {\bibfield
  {journal} {\bibinfo  {journal} {PoS}\ }\textbf {\bibinfo {volume}
  {LATTICE2012}},\ \bibinfo {pages} {010} (\bibinfo {year} {2012})},\ \Eprint
  {https://arxiv.org/abs/1210.5510} {arXiv:1210.5510 [hep-lat]} \BibitemShut
  {NoStop}%
\bibitem [{\citenamefont {Sun}\ and\ \citenamefont {Ko}(2018)}]{Sun:2018idn}%
  \BibitemOpen
  \bibfield  {author} {\bibinfo {author} {\bibfnamefont {Y.}~\bibnamefont
  {Sun}}\ and\ \bibinfo {author} {\bibfnamefont {C.~M.}\ \bibnamefont {Ko}},\
  }\href {https://doi.org/10.1103/PhysRevC.98.014911} {\bibfield  {journal}
  {\bibinfo  {journal} {Phys. Rev. C}\ }\textbf {\bibinfo {volume} {98}},\
  \bibinfo {pages} {014911} (\bibinfo {year} {2018})},\ \Eprint
  {https://arxiv.org/abs/1803.06043} {arXiv:1803.06043 [nucl-th]} \BibitemShut
  {NoStop}%
\bibitem [{\citenamefont {Kaminski}\ \emph {et~al.}(2010)\citenamefont
  {Kaminski}, \citenamefont {Landsteiner}, \citenamefont {Pena-Benitez},
  \citenamefont {Erdmenger}, \citenamefont {Greubel},\ and\ \citenamefont
  {Kerner}}]{Kaminski:2009ce}%
  \BibitemOpen
  \bibfield  {author} {\bibinfo {author} {\bibfnamefont {M.}~\bibnamefont
  {Kaminski}}, \bibinfo {author} {\bibfnamefont {K.}~\bibnamefont
  {Landsteiner}}, \bibinfo {author} {\bibfnamefont {F.}~\bibnamefont
  {Pena-Benitez}}, \bibinfo {author} {\bibfnamefont {J.}~\bibnamefont
  {Erdmenger}}, \bibinfo {author} {\bibfnamefont {C.}~\bibnamefont {Greubel}},\
  and\ \bibinfo {author} {\bibfnamefont {P.}~\bibnamefont {Kerner}},\ }\href
  {https://doi.org/10.1007/JHEP03(2010)117} {\bibfield  {journal} {\bibinfo
  {journal} {JHEP}\ }\textbf {\bibinfo {volume} {03}},\ \bibinfo {pages}
  {117}},\ \Eprint {https://arxiv.org/abs/0911.3544} {arXiv:0911.3544 [hep-th]}
  \BibitemShut {NoStop}%
\bibitem [{\citenamefont {Janiszewski}\ and\ \citenamefont
  {Kaminski}(2016)}]{Janiszewski:2015ura}%
  \BibitemOpen
  \bibfield  {author} {\bibinfo {author} {\bibfnamefont {S.}~\bibnamefont
  {Janiszewski}}\ and\ \bibinfo {author} {\bibfnamefont {M.}~\bibnamefont
  {Kaminski}},\ }\href {https://doi.org/10.1103/PhysRevD.93.025006} {\bibfield
  {journal} {\bibinfo  {journal} {Phys. Rev.}\ }\textbf {\bibinfo {volume}
  {D93}},\ \bibinfo {pages} {025006} (\bibinfo {year} {2016})},\ \Eprint
  {https://arxiv.org/abs/1508.06993} {arXiv:1508.06993 [hep-th]} \BibitemShut
  {NoStop}%
\bibitem [{\citenamefont {Kharzeev}\ \emph {et~al.}(2002)\citenamefont
  {Kharzeev}, \citenamefont {Krasnitz},\ and\ \citenamefont
  {Venugopalan}}]{Kharzeev:2001ev}%
  \BibitemOpen
  \bibfield  {author} {\bibinfo {author} {\bibfnamefont {D.}~\bibnamefont
  {Kharzeev}}, \bibinfo {author} {\bibfnamefont {A.}~\bibnamefont {Krasnitz}},\
  and\ \bibinfo {author} {\bibfnamefont {R.}~\bibnamefont {Venugopalan}},\
  }\href {https://doi.org/10.1016/S0370-2693(02)02630-8} {\bibfield  {journal}
  {\bibinfo  {journal} {Phys. Lett. B}\ }\textbf {\bibinfo {volume} {545}},\
  \bibinfo {pages} {298} (\bibinfo {year} {2002})},\ \Eprint
  {https://arxiv.org/abs/hep-ph/0109253} {arXiv:hep-ph/0109253} \BibitemShut
  {NoStop}%
\bibitem [{\citenamefont {Ikeda}\ \emph {et~al.}(2021)\citenamefont {Ikeda},
  \citenamefont {Kharzeev},\ and\ \citenamefont {Kikuchi}}]{Ikeda:2020agk}%
  \BibitemOpen
  \bibfield  {author} {\bibinfo {author} {\bibfnamefont {K.}~\bibnamefont
  {Ikeda}}, \bibinfo {author} {\bibfnamefont {D.~E.}\ \bibnamefont
  {Kharzeev}},\ and\ \bibinfo {author} {\bibfnamefont {Y.}~\bibnamefont
  {Kikuchi}},\ }\href {https://doi.org/10.1103/PhysRevD.103.L071502} {\bibfield
   {journal} {\bibinfo  {journal} {Phys. Rev. D}\ }\textbf {\bibinfo {volume}
  {103}},\ \bibinfo {pages} {L071502} (\bibinfo {year} {2021})},\ \Eprint
  {https://arxiv.org/abs/2012.02926} {arXiv:2012.02926 [hep-ph]} \BibitemShut
  {NoStop}%
\bibitem [{\citenamefont {Skokov}\ \emph {et~al.}(2009)\citenamefont {Skokov},
  \citenamefont {Illarionov},\ and\ \citenamefont {Toneev}}]{Skokov:2009qp}%
  \BibitemOpen
  \bibfield  {author} {\bibinfo {author} {\bibfnamefont {V.}~\bibnamefont
  {Skokov}}, \bibinfo {author} {\bibfnamefont {A.~{\relax Yu}.}\ \bibnamefont
  {Illarionov}},\ and\ \bibinfo {author} {\bibfnamefont {V.}~\bibnamefont
  {Toneev}},\ }\href {https://doi.org/10.1142/S0217751X09047570} {\bibfield
  {journal} {\bibinfo  {journal} {Int. J. Mod. Phys.}\ }\textbf {\bibinfo
  {volume} {A24}},\ \bibinfo {pages} {5925} (\bibinfo {year} {2009})},\ \Eprint
  {https://arxiv.org/abs/0907.1396} {arXiv:0907.1396 [nucl-th]} \BibitemShut
  {NoStop}%
\bibitem [{\citenamefont {Cartwright}\ and\ \citenamefont
  {Kaminski}(2019)}]{Cartwright2019}%
  \BibitemOpen
  \bibfield  {author} {\bibinfo {author} {\bibfnamefont {C.}~\bibnamefont
  {Cartwright}}\ and\ \bibinfo {author} {\bibfnamefont {M.}~\bibnamefont
  {Kaminski}},\ }\href {https://doi.org/10.1007/JHEP09(2019)072} {\bibfield
  {journal} {\bibinfo  {journal} {JHEP}\ }\textbf {\bibinfo {volume}
  {2019}}\bibinfo  {number} { (9)},\ \bibinfo {pages} {72}}\BibitemShut
  {NoStop}%
\end{thebibliography}%

\onecolumngrid
\appendix
\section{Equations of motion}\label{append:EOM}
For completeness we here include the equations of motion in the coordinates $z=L^2/r$.
\begin{align}
  0&= z S(v,z)^2 \left(H_1'(v,z) H_2'(v,z)+H_1'(v,z)^2+H_2'(v,z)^2\right)+z e^{-H_1(v,z)} V'(v,z)^2 \nm \\
  &+6 \left(2 S'(v,z)+z S''(v,z)\right) S(v,z)\label{eq:C_S} \, ,\\
 0  &=L^6b^2 e^{H_1(v,z)} S(v,z)^2+\left(L^3 q_5-8 \alpha  b V(v,z)\right)^2 -24 L^6 z^2 S(v,z)^4 S'(v,z) \dot{S}(v,z) \nm \\
 &-12 L^6 z^2 S(v,z)^5 \dot{S}'(v,z)-24 L^6 S(v,z)^6 \label{eq:C_Sdot} \, ,\\
 0  &= -64 \alpha ^2 b^2 e^{H_1(v,z)} V(v,z)+8 \alpha  b L^3 q_5 e^{H_1(v,z)} -L^6 z^2S(v,z)^3 \left(S'(v,z) \dot{V}(v,z)+\dot{S}(v,z) V'(v,z)\right)\nm  \\
 &+L^6 z^2 S(v,z)^4 \left(H_1'(v,z) \dot{V}(v,z)+\dot{H_1}(v,z) V'(v,z)-2 \dot{V}'(v,z)\right)\label{eq:C_Vdot}  \, ,\\
 0&=-9 z^2 S(v,z)^3 \left(H_1'(v,z) \dot{S}(v,z)+\dot{H_1}(v,z) S'(v,z)\right)-4 z^2 e^{-H_1(v,z)} S(v,z)^2 V'(v,z) \dot{V}(v,z)\nm \\
 &-6 z^2 \dot{H_1}'(v,z) S(v,z)^4 -2 b^2 e^{H_1(v,z)}\label{eq:C_Bdot}  \, ,\\
 0&=-6 z^2 \dot{H_2}'(v,z) S(v,z)^4+b^2 e^{H_1(v,z)}+2 z^2 e^{-H_1(v,z)} S(v,z)^2 V'(v,z) \dot{V}(v,z) \nm \\
 &-9 z^2 S(v,z)^3 \left(H_2'(v,z) \dot{S}(v,z)+\dot{H_2}(v,z) S'(v,z)\right) \label{eq:C_Hdot}  \, , \\
 0&=3 L^4 S(v,z)^6 \left(2 L^2 z^4 A''(v,z)+4  z^3 A'(v,z)-L^2 z^2 \dot{H_1}(v,z) \left(2 H_1'(v,z)+H_2'(v,z)\right)\right. \nm \\
 &\left.-L^2 z^2 H_1'(v,z) \dot{H_2}(v,z)-2 L^2 z^2 H_2'(v,z) \dot{H_2}(v,z)+8L^2\right)-5 b^2 L^6 e^{H_1(v,z)} S(v,z)^2 \nm \\
 &+2 L^6 z^2 e^{-H_1(v,z)} S(v,z)^4 \left(36 e^{H_1(v,z)} S'(v,z) \dot{S}(v,z)-V'(v,z) \dot{V}(v,z)\right) \nm \\
 &-7 \left(L^3 q_5-8 \alpha  b V(v,z)\right)^2\label{eq:C_A}  \, , \\
 0&=3 z^2 A'(v,z) S(v,z) \dot{S}(v,z)+L^2e^{-H_1(v,z)} \dot{V}(v,z)^2+ L^2\dot{H_1}(v,z) \dot{H_2}(v,z) S(v,z)^2\nm \\
 &+L^2\dot{H_1}(v,z)^2 S(v,z)^2+L^2\dot{H_2}(v,z)^2 S(v,z)^2+6L^2 S(v,z) \ddot{S}(v,z) \, .
\end{align}
In this list of equations, a prime denotes a radial derivative, i.e. $A'(v,r)=\partial_rA(v,r)$.
\section{Asymptotic expansion}\label{append:asymp}
In order to numerically evolve the Einstein equations we adopt coordinates $r=L^2/z$ placing the conformal boundary at $z=0$. 
 We can solve the coupled Einstein-Maxwell-Chern-Simons (EMCS) system order by order near the $AdS$ boundary. We do this for two reasons, first to extract the relevant information for the construction of the dual energy-momentum tensor and second two inform our choice of field redefinitions to improve the numerical accuracy of our solutions. The solution to the EMCS system near the conformal boundary at $z=0$ is given by,
\begin{align}
  A(v,z)&= a_4(v)z^2+\xi (v)^2+\frac{2 \xi (v)}{z}+\frac{1}{z^2} +\mathcal{O}(z^3)  \, ,\\
   H_1(v,z)&=h^{(1)}_4(v)z^4-\frac{6 v^2 \xi (v)^2+6 v \xi (v)+2}{9  v^3}z^3+\frac{ 2 v \xi (v)+1}{3 v^2}z^2-\frac{2z}{3  v} \nm \\
   &-\frac{2}{3}\log \left(v\right) +\mathcal{O}(z^5)  \, , \\
    H_2(v,z)&= h^{(2)}_4(v)z^4-\frac{6 v^2 \xi (v)^2+6 v \xi (v)+2}{9 v^3}z^3+\frac{ 2 v \xi (v)+1}{3 v^2}z^2-\frac{2 z}{3 v} \nm \\
    &-\frac{2 \log (v)}{3}+\mathcal{O}(z^5)  \, ,  \\
   S(v,z)&=\frac{3 v \xi (v)+1}{3 v^{2/3}}+\frac{ (9 v \xi (v)+5)}{81 v^{8/3}}z^2-\frac{1}{9 v^{5/3}}z+\frac{v^{1/3}}{z}+\mathcal{O}(z^3)  \, ,\\
   \phi(v,z)&=\frac{q_5}{2 v}z^2+\mathcal{O}(z^3) \, , \\
   V(v,z) &= V_2(v) z^2+\mathcal{O}(z^3)  \, ,
\end{align}
where we have set $L=1$ to reduce clutter in the equations. Factors of $L$ can be restored through dimensional analysis. The coefficients $a_4,h^{(1)}_4,h^{(2)}_4,V_2, q_5$ cannot be determined by a near-boundary analysis and represent either data to be extracted from our evolution $(a_4,h^{(1)}_4,h^{(2)}_4,V_2)$ or data that must be supplied to begin our evolution $(a_4(\tau_0),h^{(1)}_4(\tau_0),h^{(2)}_4(\tau_0),V_2(\tau_0),q_5)$.

\section{Numerical algorithm}\label{append:num}
\textbf{Setting Initial Data:} We begin by fixing $H_1(v_i,z;\xi_i)$ and $H_2(v_i,z;\xi_i)$ on the initial time step where $\xi_i$ is a guess for the initial radial shift. We then solve the linear equation given by eq.\,(\ref{eq:C_S}) in the limit of vanishing Chern-Simons coupling and vanishing bulk electric field $V(v,z)$, for $S_{\text{Linear}}$. Given $S_{\text{Linear}}$ we fix $V(v,z)$ on the initial time step and solve the nonlinear system for $S(v,z)$ using Fr\'{e}chet differentiation and Newton iteration. The linear solution $S_{\text{Linear}}$ serves as an initial guess for this non-linear relaxation approach. We then solve eq.\,(\ref{eq:C_Sdot}) and locate the apparent horizon by solving the equation for the roots of the expansion of a congruence of null geodesics, $\theta=\dot{S}(v_i,A_h)=0$. Once we have obtained the location of the apparent horizon we use the residual shift symmetry to place the horizon at $z=1$ by setting $\xi=\xi_i+1-1/A_h$. 

\textbf{Numerical Evolution:} After setting the initial data and obtaining $\xi$ on the initial time slice we repeat our initial step of solving the equation for $S$ by fixing $H_1(v_i,z;\xi)$ and $H_2(v_i,z;\xi)$. Once we have a solution we proceed to solve eq.\,\ (\ref{eq:C_Sdot}) for $\dot{S}$. We then solve the coupled system, eq.\,\ (\ref{eq:C_Vdot}) through (\ref{eq:C_Hdot}) for the functions $(\dot{V},\dot{H}_1,\dot{H}_2)$. We then check the location of the apparent horizon and demand the time independence of the expansion $\partial_v\theta=0$ to obtain a boundary condition for the metric function $A$. With our boundary condition we can solve eq.\ (\ref{eq:C_A}) completing our run through the "pseudo-nested" system.

After each run through the "pseudo-nested" system we extract the time derivatives $\partial_vH_1(v,z)\partial_vH_2(v,z)$ and $\partial_vV(v,z)$ from the definition in eq.~(\ref{eq:char_dot}) and use them to construct data for $H_1,H_2$ and $V$ on the next time step. The time derivative of $\xi$ is found by extracting it from the near-boundary behavior of the function $A$. By use of appropriate redefinitions this can be done such that,
\begin{equation}
    \xi'(v)=\lim_{z\rightarrow 0} -\frac{1}{2}A_s(z,v) \, .
\end{equation}
On each subsequent time slice we use the previous solution to eq.\ (\ref{eq:C_S}) as an initial guess for Newton iteration of the non-linear system. We continue to repeat the procedure until we reach the specified final time.

\end{document}